\documentclass[journal]{IEEEtran}
\usepackage{xcolor,soul,framed}
\colorlet{shadecolor}{yellow}
\usepackage{graphicx}
\graphicspath{{../pdf/}{../jpeg/}}
\DeclareGraphicsExtensions{.pdf,.jpeg,.png}
\usepackage[cmex10]{amsmath}
\usepackage{array}
\usepackage{mdwmath}
\usepackage{mdwtab}
\usepackage{eqparbox}
\usepackage{url}
\usepackage{enumerate}
\usepackage{multirow} 
\usepackage{balance}
\usepackage{amsmath}
\usepackage{algorithmic}
\usepackage{array}
\begin{document}

\title{Review of Smart Meter Data Analytics: Applications, Methodologies, and Challenges}

\author{Yi Wang,~\IEEEmembership{}
        Qixin Chen,~\IEEEmembership{}
        Tao Hong,~\IEEEmembership{}
        Chongqing Kang~\IEEEmembership{}
\thanks{Manuscript received September 2, 2017; revised January 01, 2018; accepted February 15, 2018. Date of publication March **, 2018; date of current version December **, 2019. This work was jointly supported by National Key R\&D Program of China (No. 2016YFB0900100) and the Major Smart Grid Joint Project of National Natural Science Foundation of China and State Grid (No. U1766212). Paper no. TSG-01277-2017. \textit{(Corresponding Author: Chongqing Kang})}
\thanks{Y. Wang, Q. Chen, and C. Kang are with the State Key Laboratory of Power Systems, Department. of Electrical Engineering, Tsinghua University, Beijing 100084, China. (E-mail: cqkang@tsinghua.edu.cn).}
\thanks{T. Hong is with the  Energy  Production  and  Infrastructure Center, University of North Carolina at Charlotte, Charlotte, NC 28223 USA.}
\thanks{Color  versions  of  one  or  more  of  the  figures  in  this  paper  are  available online at http://ieeexplore.ieee.org.}
\thanks{Digital Object Identifier 10.1109/TSG.2018.2805***}}
\markboth{IEEE Trans. Smart Grid, Accepted.}
{Shell \MakeLowercase{\textit{et al.}}: Bare Demo of IEEEtran.cls for IEEE Journals}
\maketitle

\begin{abstract}
The widespread popularity of smart meters enables an immense amount of fine-grained electricity consumption data to be collected. Meanwhile, the deregulation of the power industry, particularly on the delivery side, has continuously been moving forward worldwide. How to employ massive smart meter data to promote and enhance the efficiency and sustainability of the power grid is a pressing issue. To date, substantial works have been conducted on smart meter data analytics. To provide a comprehensive overview of the current research and to identify challenges for future research, this paper conducts an application-oriented review of smart meter data analytics. Following the three stages of analytics, namely, descriptive, predictive and prescriptive analytics, we identify the key application areas as load analysis, load forecasting, and load management. We also review the techniques and methodologies adopted or developed to address each application. In addition, we also discuss some research trends, such as big data issues, novel machine learning technologies, new business models, the transition of energy systems, and data privacy and security.
\end{abstract}


\IEEEpeerreviewmaketitle
\section{Introduction}
\IEEEPARstart{S}{mart} meters have been deployed around the globe during the past decade. For example, the numbers of smart meters installed in the UK, the US, and China reached 2.9 million \cite{UKSmartmeter}, 70 million \cite{USSmartmeter1}\cite{USSmartmeter2}, and 96 million, respectively by the end of 2016. 
Smart meters, together with the communication network and data management system, constitute the advanced metering infrastructure (AMI), which plays a vital role in power delivery systems by recording the load profiles and facilitating bi-directional information flow \cite{mohassel2014survey}. The widespread popularity of smart meters enables an immense amount of fine-grained electricity consumption data to be collected. Billing is no longer the only function of smart meters. High-resolution data from smart meters provide rich information on the electricity consumption behaviors and lifestyles of the consumers. 
Meanwhile, the deregulation of the power industry, particularly on the delivery side, is continuously moving forward in many countries worldwide. These countries are now sparing no effort on electricity retail market reform. Increasingly more participators, including retailers, consumers, and aggregators, are involved in making the retail market more prosperous, active, and competitive \cite{yang2017decision}. How to employ massive smart meter data to promote and enhance the efficiency and sustainability of the demand side has become an important topic worldwide. 

In recent years, the power industry has witnessed considerable developments of data analytics in the processes of generation, transmission, equipment, and consumption. 
Increasingly more projects on smart meter data analytics have also been established.
The National Science Foundation (NSF) of the United States provides a standard grant for cross-disciplinary research on smart grid big data analytics \cite{NSF}. Several projects for smart meter data analytics are supported by the CITIES Innovation Center in Denmark. These projects investigate machine learning techniques for smart meter data to improve forecasting and money-saving opportunities for customers \cite{liu2017citiesdata}. The Bits to Energy Lab which is a joint research initiative of ETH Zurich, the University of Bamberg, and the University of St. Gallen, has launched several projects for smart meter data analytics for customer segmentation and scalable efficiency services \cite{Bit2Energy}. The Siebel Energy Institute, a global consortium of innovative and collaborative energy research, funds cooperative and innovative research grants for data analytics in smart girds \cite{Siebel}.  Meanwhile, the National Science Foundation of China (NSFC) and the National Key R\&D Program of China are approving increasingly more data-analytics-related projects in the smart grid field, such as the National High Technology Research and Development Program of China (863 Program) titled Key Technologies of Big Data Analytics for Intelligent Distribution and Utilization. ESSnet Big Data, a project within the European statistical system (ESS), aims to explore big data applications, including smart meters \cite{WP3}. The workpackage in the ESSnet Big Data project concentrates on smart meter data access, handling, and deployments of methodologies and techniques for smart meter data analytics. National statistical institutes from Austria, Denmark, Estonia, Sweden, Italy, and Portugal jointly conduct this project.

Apart from academic research, data analytics has already been used in industry. In June 2017, SAS published the results from its industrial analytics survey \cite{SAS}. This survey aims to provide the issues and trends shaping how utilities deploy data and analytics to achieve business goals. There are 136 utilities from 24 countries that responded to the survey. The results indicate that data analytics application areas include energy forecasting, smart meter analytics, asset management/analytics, grid operation, customer segmentation, energy trading, credit and collection, call center analytics, and energy efficiency and demand response program engagement and marketing. More and more energy data scientists will be jointly trained by universities and industry bridge the talent gap in energy data analytics \cite{PEM4}. Meanwhile, the privilege of smart meters and deregulation of the demand side are accelerating the birth of many start-ups. These start-ups attempt to collect and analyze smart meter data and provide insights and value-added services for consumers and retailers to make profits. More details regarding industrial applications can be found from the businesses of the data-analytics-based start-ups. 

Analytics is known as the scientific process of transforming data into insights for making better decisions. It is commonly dissected into three stages: descriptive analytics (what do the data look like), predictive analytics (what is going to happen with the data), and prescriptive analytics (what decisions can be made from the data). This review of smart meter data analytics is conducted from these three aspects.

\subsection{Bibliometric analysis}
To provide an overview of the existing research in smart meter data analytics, a bibliometric analysis was conducted on 31 December 2017 using the well-established and acknowledged databases, Web of Science (WoS). The query for WoS is as follows: TS=(("smart meter" OR "consumption" OR "demand" OR "load") AND "data" AND ("household" OR "resident" OR "residential" OR "building" OR "industrial" OR "individual" OR "customer" OR "consumer") AND ("energy theft" OR "demand response" OR "clustering" OR "forecasting" OR "profiling" OR "classification" OR "abnormal" OR "anomaly") AND ("smart grid" OR "power system")).

\begin{figure}[!t]
  \begin{center}
  \includegraphics[width=3in]{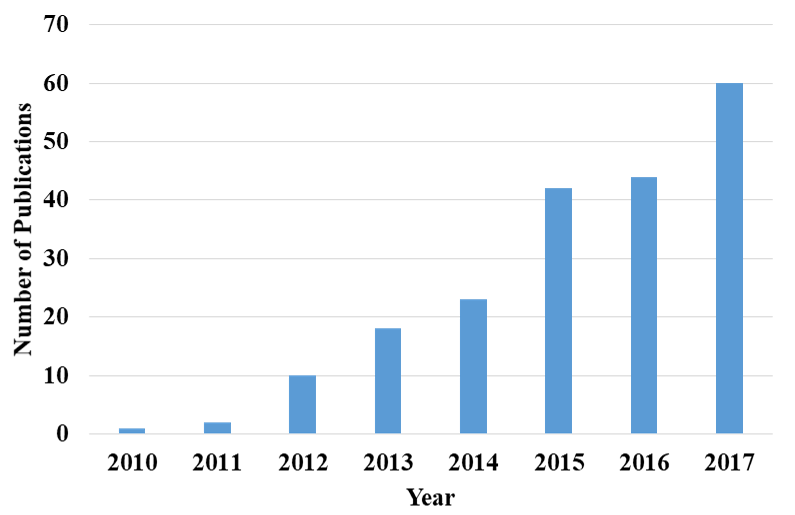}\\
  \caption{Number of publications indexed by WoS.}\label{IndexPublications}
  \end{center}
\end{figure}

Fig. \ref{IndexPublications} shows the number of publications indexed by WoS 2010 to 2017. In total, 200 publications were found in WoS. 
Before 2011, the number of publications was at a relatively low level, while it increased rapidly beginning in 2012 and reached 60 in WoS, in the year of 2017. This result should not be a surprise. The smart grid initiatives started around the late 2000s. It takes a few years for power companies to collect the data for extensive research and another few years to bring the research findings to journal publications.

\begin{figure}[!t]
  \begin{center}
  \includegraphics[width=3.5in]{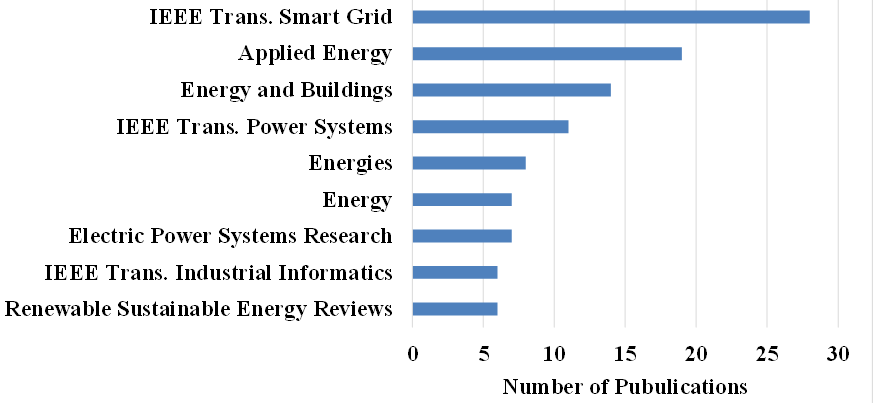}\\
  \caption{Number of publications in nine most popular journals.}\label{JournalPublications}
  \end{center}
\end{figure}

Fig. \ref{JournalPublications} depicts the journals ranked by the number of relevant papers published since 2010 according to WoS. IEEE Transactions on Smart Grid, the youngest journal on this list, has published 28 relevant papers since it was founded in 2012, making it the most popular journal for smart meter data analytics articles. Among the top 5 most popular journals, Energy and Buildings is not a traditional venue for power engineering papers since smart meters are tied to residential homes and office buildings, which makes this journal a natural outlet for smart meter data analytics papers. 

\subsection{Relevant Review Articles}
Here, we conduct a brief review of several review articles related to smart meter data analytics. Note that these existing reviews are either beyond the scope of this paper or not as comprehensive and up-to-date as this paper. 

The challenges and opportunities related to the privacy and security aspects of energy data analytics were discussed in \cite{hu2016energy}. Several reviews of load profiling have been conducted from the aspects of clustering method comparison \cite{chicco2012overview}, applications \cite{zhou2013review} and demand response \cite{wang2015load}. A few review articles focused on non-intrusive load monitoring (NILM), such as different disaggregation algorithms \cite{zeifman2011nonintrusive} and the evaluation criteria \cite{esa2016review}. The methods for how to detect non-technical loss (NTL) in power grids based on smart meter data were reviewed in \cite{ahmad2017non}, while the NTL challenges were discussed in \cite{glauner2017challenge}. The applications of smart meter data analytics on customer behavior were summarized in \cite{chicco2016customer}. Load forecasting is the main application of smart meter data analytics. A natural application of smart meter data is to enhance load forecast accuracy. The recent developments of point and probabilistic load forecasting were critically reviewed in \cite{hong2016probabilistic}. Several forecasting techniques for building loads were summarized in \cite{deb2017review}. A comprehensive review of smart meter analytics was conducted in \cite{alahakoon2016smart}, which covered several large areas, including the smart metering environments, analysis techniques, and potential applications. Our paper differs from Ref. \cite{alahakoon2016smart} by only focusing on data analytics applications and methodologies rather than including data collection and communication.
The review papers mentioned above summarized the application of smart meter data analytics from a specific aspect. In addition, under the background of the big data era and further opening of the retail market, smart meter data analytics is still an emerging research field. More data analytics methods have been studied, and more novel problems have been defined. This paper attempts to provide a comprehensive review of smart meter data analytics in wider applications. More emphasis is placed on the research over the past five years. Classical and typical works of literature that were published earlier are also included.

\subsection{Open Load Datasets}
Due to many issues, such as privacy and security, many power companies are hesitant to release their smart meter data to the public. This has been a challenge for conducting research in smart meter data analytics and its applications. Nevertheless, several anonymized or semi-anonymized datasets at the household level have been made publicly available over the past few years. Various studies have been conducted based on these smart meter datasets. Several open load datasets are summarized in Table \ref{OpenDataset}.

\begin{table*}[]
\centering
\caption{Basic Information of Several Open Load Datasets}
\label{OpenDataset}
\begin{tabular}{|c|l|c|c|c|c|}
\hline
Name                     & \multicolumn{1}{c|}{Brief Description}                                                                                               & Number & Frequency    & Duration & References\\ \hline
Customer Behavior Trials \cite{Irish} & \begin{tabular}[c]{@{}l@{}}Smart meter read data;\\ Pre- and  post-trial survey data;\end{tabular}                 & 6445            & Every 30 min & 2009/9-2011/1 &
\begin{tabular}[c]{@{}l@{}}\cite{jokar2016electricity}\cite{wang2016sparse}\cite{shi2017deep}\cite{wang2016clustering}\\ \cite{chaouch2014clustering}\cite{quilumba2015using}\cite{taieb2016forecasting}\cite{tong2016cross}\\ \cite{mcloughlin2015clustering}\cite{beckel2014revealing}\cite{tong2016smart}\cite{8291011}\end{tabular}
\\  \hline
Low Carbon London \cite{schofield2015low}  & \begin{tabular}[c]{@{}l@{}}Smart meter read data;\\ Electricity price data;\\ Appliance and attitude survey data;\end{tabular} & 5567            & Every 30 min & 2013/1-2013/12 &\cite{sun2016c}\cite{mingyang} \\ \hline
PecanStreet \cite{PecanStreet}              & \begin{tabular}[c]{@{}l@{}}Residential electricity consumption data;\\ Electric vehicle charging data;\end{tabular}                                                                                                                     & 500             & Every 1 min & 2005/5-2017/5&   \\ \hline
Building Data Genome \cite{MILLER2017439}              & \begin{tabular}[c]{@{}l@{}}Non-residential building smart meter data;\\ Area, weather, and primary use type data;\end{tabular}                                                                                                                     & 507             & Every 1 hour & 2014/12-2015/11&   \\ \hline
UMass Smart \cite{UMass}              & Residential electricity consumption data;                                                                                                                    & 400             & Every 1 min & One day&\cite{marnerides2015power}\cite{7947198}   \\ \hline
Ausgrid Residents \cite{ratnam2017residential}        & \begin{tabular}[c]{@{}l@{}}General consumption data;\\ Controlled load consumption data;\\ PV output data;\end{tabular}        & 300             & Every 30 min & 2010/7-2013/6 &  \\ \hline
Ausgrid Substation \cite{AusSubstation}       & Substation metering data;                                                                                                      & 177             & Every 15 min & 2005/5- & \\ \hline
GEFCom 2012 \cite{Hong2014Global}          & \begin{tabular}[c]{@{}l@{}}Zonal load data;\\ Temperature data;\end{tabular}                                                 & 20              &  Hourly            &  2003/1-2008/6 &\begin{tabular}[c]{@{}l@{}}\cite{hoverstad2015short}\cite{liu2015probabilistic}\cite{charlton2014refined}\cite{thouvenot2016electricity}\\\cite{lloyd2014gefcom2012}\cite{nedellec2014gefcom2012}\cite{taieb2014gradient}\cite{hong2015weather} \end{tabular}
\\ \hline
GEFCom 2014 \cite{hong2016probabilistic123}          & \begin{tabular}[c]{@{}l@{}}Zonal load data;\\ 25 weather station data;\end{tabular}                                                 & 1              &  Hourly            &  2005/1-2010/9 &\begin{tabular}[c]{@{}l@{}}\cite{antoniadis2016prediction}\cite{dordonnat2016gefcom2014}\cite{xie2016gefcom2014}\cite{haben2016hybrid}\\\cite{mangalova2016sequence}\cite{ziel2016lasso}\cite{gaillard2016additive}\end{tabular}           \\ \hline
ISO New England \cite{ISONewEngland}          & \begin{tabular}[c]{@{}l@{}}System load data;\\ Temperature data;\\ Locational marginal pricing data;\end{tabular}                                                 & 9              &  Hourly            &  2003/1-& \cite{wang2016electric}\cite{xie2016relative}\cite{xie2017wind}\cite{xie2016temperature}             \\ \hline
\end{tabular}
\end{table*}

\begin{itemize}
\item \textbf{Customer Behavior Trials}: The Commission for Energy Regulation (CER), the regulator for the electricity and natural gas sectors in Ireland, launched a smart metering project to conduct customer behavior trials (CBTs) to determine how smart metering can help shape energy usage behaviors across a variety of demographics, lifestyles, and home sizes \cite{Irish}.
\item \textbf{Low Carbon London}: Similar to CBTs, a trial as a part of the Low Carbon London project involved over five thousand households in the London area \cite{schofield2015low}. Smart meter data, time-of-use tariff data, and survey data were collected to investigate the impacts of a wide range of low carbon technologies on London's electricity distribution network.
\item \textbf{PecanStreet}: The dataset is supported by the Pecan Street experiment in Austin, TX. The dataset contains minute-level electricity consumption data from 500 homes (both the whole-home level and individually monitored appliance circuits) \cite{PecanStreet}.
\item \textbf{Building Data Genome}: The dataset is from The Building Data Genome Project. There are 507 whole-building electrical meters in this collection, and the majority are from buildings on university campuses \cite{MILLER2017439}.
\item \textbf{UMass Smart}: This dataset was collected by the Mass Smart Microgrid project, which contains consumption data of 400 homes at the one-minute granularity of a whole day \cite{UMass}.
\item \textbf{Ausgrid Resident}: The smart meter data combined with rooftop PV generation data of 300 residential consumers in an Australian distribution network over 3 years have been recorded by the Ausgrid distribution network \cite{ratnam2017residential}.
\item \textbf{Ausgrid Substation}: Ausgrid also makes the load profiles of approximately 180 zone substations publicly available from 2005 \cite{AusSubstation}. The dataset is continuously updated.
\item \textbf{GEFCom2012}: About four and a half years (Janurary 2004 to July 2008) of hourly load and temperature data were provided. The temperature readings were from 11 weather stations. The top ranked methods in the competition were summarized in \cite{Hong2014Global}.
\item \textbf{GEFCom2014}: The dataset contains the hourly load data and 25 weather station data during 2005 and 2010. The detailed description of the dataset and many practical and high-rank forecasting methods in this competition are summarized in \cite{hong2016probabilistic123}.
\item \textbf{ISO New England}: ISO New England publishes the system-level load data and corresponding temperature data of 9 zones every month. More information can be found in \cite{ISONewEngland}.
\end{itemize}

As shown, the datasets including Low Carbon London, Ausgrid Residents, and Ausgrid Substation have seldom been used in the existing literature. In addition, Building Data Genome is a newly released dataset. More works can be performed based on these datasets.

\subsection{Taxonomy}
\label{Taxonomy}
\begin{figure*}[!t]
  \begin{center}
  \includegraphics[width=6in]{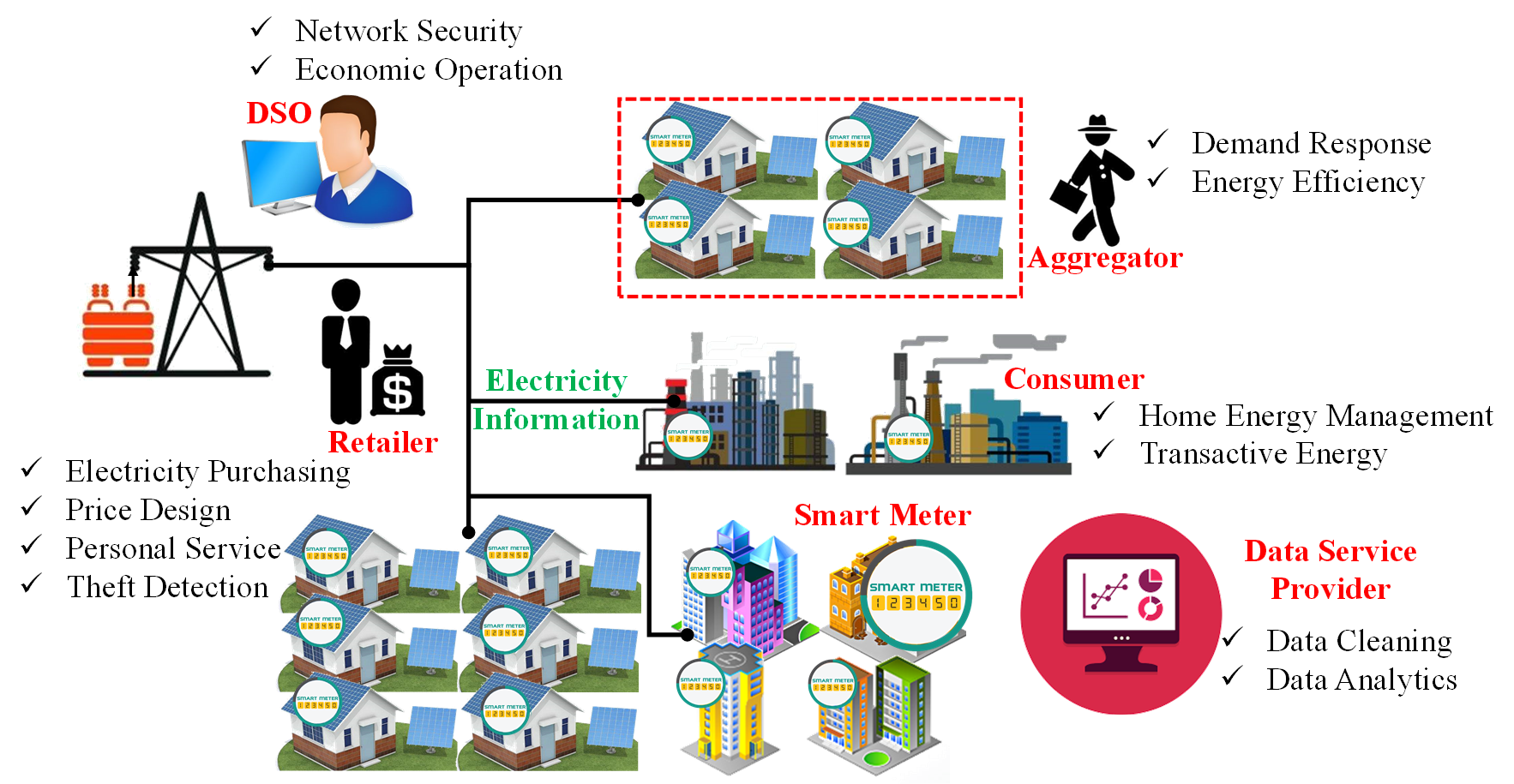}\\
  \caption{Participators and their businesses on the demand side.}\label{Retail_Market}
  \end{center}
\end{figure*}

Fig. \ref{Retail_Market} depicts the five major players on the demand side of the power system: consumers, retailers, aggregators, distribution system operators (DSO), and data service providers. 

For retailers, at least four businesses related to smart meter data analytics need to be conducted to increase the competitiveness in the retail market. 1) Load forecasting, which is the basis of decision making for the optimization of electricity purchasing in different markets to maximize profits. 2) Price design to attract more consumers. 3) Providing good service to consumers, which can be implemented by consumer segmentation and characterization. 4) Abnormal detection to have a cleaner dataset for further analysis and decrease potential loss from electricity theft. For consumers, individual load forecasting, which is the input of future home energy management systems (HEMS) \cite{7745930}, can be conducted to reduce their electricity bill. In the future peer-to-peer (P2P) market, individual load forecasting can also contribute to the implementation of transactive energy between consumers \cite{pratt2016transactive}\cite{morstyn2018using}. 
For aggregators, they deputize a group of consumers for demand response or energy efficiency in the ancillary market. Aggregation level load forecasting and demand response potential evaluation techniques should be developed. For DSO, smart meter data can be applied to distribution network topology identification, optimal distribution system energy management, outage management, and so forth. For data service providers, they need to collect  smart meter data and then analyze these massive data and provide valuable information for retailers and consumers to maximize profits or minimize cost. Providing data services including data management, data analytics is an important business model when increasingly more smart meter data are collected and to be processed.

\begin{figure*}[!t]
  \begin{center}
  \includegraphics[width=6in]{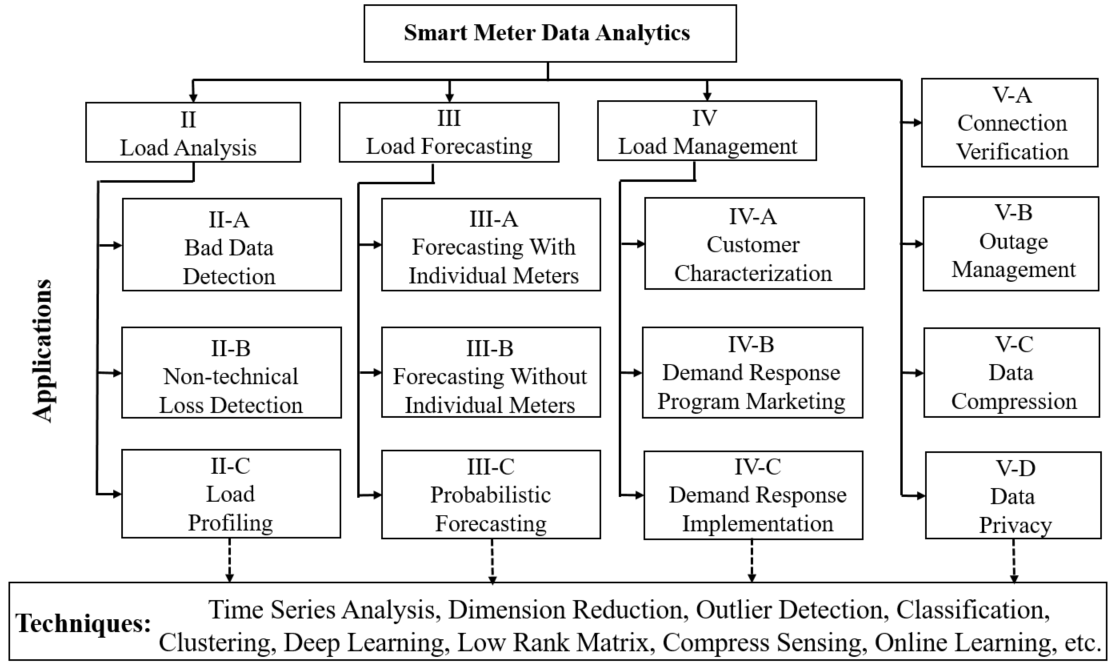}\\
  \caption{Taxonomy of smart meter data analytics.}\label{Framework}
  \end{center}
\end{figure*}

To support the businesses of retailers, consumers, aggregators, DSO, and data service providers, following the three stages of analytics, namely, descriptive, predictive and prescriptive analytics, the main applications of smart meter data analytics are classified into load analysis, load forecasting, load managements, and so forth. The detailed taxonomy is illustrated in Fig. \ref{Framework}. The main machine learning techniques used for smart meter data analytics include time series, dimensionality reduction, clustering, classification, outlier detection, deep learning, low-rank matrix, compressed sensing, online learning, and so on. Studies on how smart meter data analytics works for each application and what methodologies have been applied will be summarized in the following sections.

\subsection{Contributions and Organization}
This paper attempts to provide a comprehensive review of the current research in recent years and identify future challenges for smart meter data analytics. Note that every second or higher frequency data used for NILM are very limited at present due to the high cost of communicating and storing the data. The majority of smart meters collect electricity consumption data at a frequency of every 15 minutes to each hour. In addition, several comprehensive reviews have been conducted on NILM. Thus, in this review paper, works about NILM are not included.

The contributions of this paper are as follows:
  \begin{enumerate}
    \item Conducting a comprehensive literature review of smart meter data analytics on the demand side with the newest developments, particularly over the past five years.
    \item Providing a well-designed taxonomy for smart meter data analytics applications from the perspective of load analysis, load forecasting, load management, and so forth.
    \item Discussing open research questions for future research directions, including big data issues, new machine learning technologies, new business models, the transition of energy systems, and data privacy and security.
  \end{enumerate}

The remainder of the paper is organized as follows. Sections \ref{abnormal} to \ref{demandresponse} conduct the survey on smart meter data analytics for load analysis, load forecasting, and load management, respectively. Section \ref{miscellanies} provides a miscellany of smart meter data analytics in addition to the three aspects above. Section \ref{openissue} discusses several future research issues. Section \ref{conclusion} draws the conclusions.

\section{Load Analysis}
\label{abnormal}
\begin{figure}[!t]
  \begin{center}
  \includegraphics[width=3.5in]{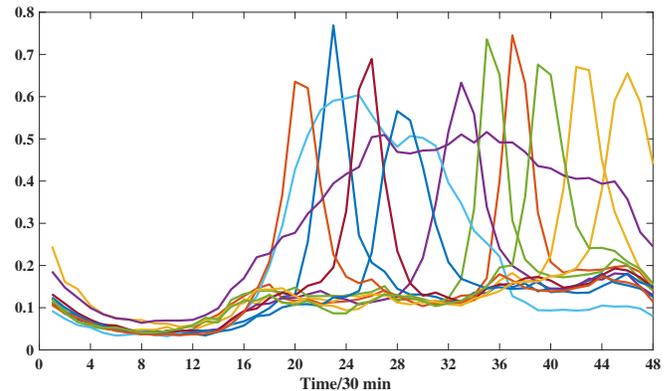}\\
  \caption{Typical normalized daily residential load profiles.}\label{Resident}
  \end{center}
\end{figure}

Fig. \ref{Resident} shows eight typical normalized daily residential load profiles obtained using the simple k-means algorithm in the Irish resident load dataset. The load profiles of different consumers on different days are diverse. Having a better understanding of the volatility and uncertainty of the massive load profiles is very important for further load analysis.
In this section, the works on load analysis are reviewed from the perspectives of anomaly detection and load profiling. Anomaly detection is very important because training a model such as a forecasting model or clustering model on a smart meter dataset with anomalous data may result in bias or failure for parameter estimation and model establishment. Moreover, reliable smart meter data are important for accurate billing. The works on anomaly detection in smart meter data are summarized from the perspective of bad data detection and NTL detection (or energy theft detection). Load profiling is used to find the basic electricity consumption patterns of each consumer or a group of consumers. The load profiling results can be further used for load forecasting and demand response programs. 

\subsection{Bad Data Detection}
Bad data as discussed here can be missing data or unusual patterns caused by unplanned events or the failure of data collection, communication, or entry. Bad data detection can be divided into probabilistic, statistical, and machine learning methods \cite{hodge2004survey}. The methods for bad data detection in other areas can be directly applied to smart meter data. Only the works closely related to smart meter bad data detection are surveyed in this subsection. According to the modeling methods, these works are summarized as time-series-based methods, low-rank matrix technique-based methods, and time-window-based methods. 

Smart meter data are essentially time series. An optimally weighted average (OWA) method was proposed for data cleaning and imputation in \cite{peppanen2016handling}, which can be applied to offline or online situations. It was assumed that the load data can be explained by a linear combination of the nearest neighbor data, which is quite similar to the autoregressive moving average (ARIMA) model for time series. The optimal weight was obtained by training an optimization model. While in \cite{akouemo2017data}, the nonlinear relationship between the data at different time periods and exogenous inputs was modeled by combining autoregressive with exogenous inputs (ARX) and artificial neural network (ANN) models where the bad data detection was modeled as a hypothesis testing on the extreme of the residuals. A case study on gas flow data was performed and showed an improvement in load forecasting accuracy after ARX-based bad data detection. Similarly, based on the auto-regression (AR) model, the generalized extreme Studentized deviate (GESD) and the \textit{Q}-test were proposed to detect the outliers when the number of samples is more and less than ten, respectively, in \cite{li2010classification}. Then, canonical variate analysis (CVA) was conducted to cluster the recovered load profiles, and a linear discriminate analysis (LDA) classifier was further used to search for abnormal electricity consumption. Instead of detecting bad data, which forecasting method is robust to the cyber attack or bad data without bad data detection was investigated in \cite{taoMPCE}. 

The electricity consumptions are spatially and temporally correlated. Exploring the spatiotemporal correlation can help identify the outliers and recover them. A low-rank matrix fitting-based method was proposed in \cite{mateos2013load} to conduct data cleaning and imputation. An alternating direction method of multipliers (ADMM)-based distributed low-rank matrix technique was also proposed to enable communication and data exchange between different consumers and to protect the privacy of consumers. Similarly, to produce a reliable state estimation, the measurements were first processed by low-rank matrix techniques in \cite{huang2017false}. Both off-line and on-line algorithms have been proposed. However, the improvement in state estimation after low-rank denoising has not been investigated. Low-rank matrix factorization works well when the bad data are randomly distributed. However, when the data are unchanged for a certain period, the low-rank matrix cannot handle it well. More data preparation works were conducted to detect these types of bad data before singular value thresholding (SVT)-based low-rank matrix-based bad data identification and recovery in \cite{wang2016application}. 

Rather than detecting all the bad data directly, strategies that continuously detect and recover a part within a certain time window have also been studied. A clustering approach was proposed on the load profiles with missing data in \cite{al2016k} and \cite{al2016state}. The clustering was conducted on segmented profiles rather than the entire load profiles in a rolling manner. In this way, the missing data can be recovered or estimated by other data in the same cluster. A collective contextual anomaly detection using a sliding window framework was proposed in \cite{araya2017ensemble} by combining various anomaly classifiers. The anomalous data were detected using overlapping sliding windows. Since smart meter data are collected in a real-time or near real-time fashion, an online anomaly detection method using the Lambda architecture was proposed in \cite{liu2016online}. The proposed online detection method can be parallel processed, having high efficiency when working with large datasets.

\subsection{Energy Theft Detection}
Strictly speaking, smart meter data with energy theft also belong to bad data. The bad data discussed above are unintentional and appear temporarily, whereas energy theft may change the smart meter data under certain strategies and last for a relatively long time. Energy theft detection can be implemented using smart meter data and power system state data, such as node voltages. The energy theft detection methods with only smart meter data are summarized in this part from two aspects: supervised learning and unsupervised learning.

Supervised classification methods are effective approaches for energy theft detection, which generally consists of two stages: feature extraction and classification. 
To train a theft detection classifier, the non-technical loss was first estimated in \cite{jokar2016electricity}. K-means clustering was used to group the load profiles, where the number of clusters was determined by the silhouette value \cite{wang2009cvap}. To address the challenge of imbalanced data, various possible malicious samples were generated to train the classifier. An energy theft alarm was raised after a certain number of abnormal detections. Different numbers of abnormal detections resulted in different false positive rates (FPR) and Bayesian detection rates (BDR). The proposed method can also identify the energy theft types. Apart from clustering-based feature extraction, an encoding technique was first performed on the load data in \cite{depuru2013high}, which served as the inputs of classifiers including SVM and a rule-engine-based algorithm to detect the energy theft. The proposed method can run in parallel for real-time detection. By introducing external variables, a top-down scheme based on decision tree and SVM methods was proposed in \cite{jindal2016decision}. The decision tree estimated the expected electricity consumption based on the number of appliances, persons, and outdoor temperature. Then, the output of the decision tree was fed to the SVM to determine whether the consumer is normal or malicious. The proposed framework can also be applied for real-time detection. 

Obtaining the labeled dataset for energy theft detection is difficult and expensive. Compared with supervised learning, unsupervised energy theft detection does not need the labels of all or partial consumers. An optimum-path forest (OPF) clustering algorithm was proposed in \cite{junior2016unsupervised}, where each cluster is modeled as a Gaussian distribution. The load profile can be identified as an anomaly if the distance is greater than a threshold. Comparisons with frequently used methods, including k-means, Birch, affinity propagation (AP), and Gaussian mixture model (GMM), verified the superiority of the proposed method. Rather than clustering all load profiles, clustering was only conducted within an individual consumer to obtain the typical and atypical load profiles in \cite{nizar2008power}. A classifier was then trained based on the typical and atypical load profiles for energy theft detection. A case study in this paper showed that extreme learning machine (ELM) and online sequential-ELM (OS-ELM)-based classifiers have better accuracy compared with SVM. Transforming the time series smart meter data into the frequency domain is another approach for feature extraction. Based on the discrete Fourier transform (DFT) results, the features extracted in the reference interval and examined interval were compared based on the so-called \textit{Structure} \& \textit{Detect} method in \cite{botev2016detecting}. Then, the load profile can be determined to be a normal or malicious one. The proposed method can be implemented in a parallel and distributed manner, which can be used for the on-line analysis of large datasets. Another unsupervised energy theft detection method is to formulate the problem as a load forecasting problem. If the metered consumption is considerably lower than the forecasted consumption, then the consumer can be marked as a malicious consumer. An anomaly score was given to each consumption data and shown with different colors to realize visualization in \cite{janetzko2014anomaly}.  

\subsection{Load Profiling}
Load profiling refers to the classification of load curves or consumers according to electricity consumption behaviors. In this subsection, load profiling is divided into direct-clustering-based and indirect-clustering-based approaches. Various clustering techniques, such as K-means, hierarchical clustering, and self-organizing map (SOM), have been directly implemented on smart meter data \cite{chicco2012overview}\cite{zhou2013review}\cite{wang2015load}. Two basic issues about direct clustering are first discussed. Then, the works on indirect clustering are classified into dimensionality reduction, load characteristics, and variability and uncertainty-based methods according to the features that are extracted before clustering.

There are some basic issues associated with direct clustering. The first issue is the resolution of smart meter data. In \cite{granell2015impacts}, three frequently used clustering techniques, namely, k-means, hierarchical algorithms, and the Dirichlet process mixture model (DPMM) algorithm, were performed on the smart meter data with different frequencies varying from every 1 minute to 2 hours to investigate how the resolution of smart meter data influences the clustering results. The results showed that the smart meter data with a frequency of at least every 30 minutes is sufficiently reliable for most purposes. The second issue is that the smart meter data are essentially time-series data.  In contrast to traditional clustering methods for static data, k-means modified for dynamic clustering was proposed in \cite{benitez2014dynamic} to address time-dependent data. The dynamic clustering allows capturing the trend of clusters of consumers. A two-stage clustering strategy was proposed in \cite{7947198} to reduce the computational complexity. In the first stage, K-means was performed to generate the local representative load profiles; in the second stage, clustering was further performed on the clustering centers obtained in the first stage at the central processor. In this way, the clustering method can be performed in a distributed fashion and largely reduce the overall complexity.

Apart from direct clustering, increasingly more literatures are focusing on indirect clustering, i.e., feature extraction is conducted before clustering. Dimensionality reduction is an effective way to address the high dimensionality of smart meter data. Principal component analysis (PCA) was performed on yearly load profiles to reduce the dimensionality of original data and then k-means was used to classify consumers in \cite{koivisto2013clustering}. The components learned by PCA can reveal the consumption behaviors of different connection point types. Similarly, PCA was also used to find the temporal patterns of each consumer and spatial patterns of several consumers in \cite{chelmis2015big}. Then, a modified K-medoids algorithm based on the Hausdorff distance and Voronoi decomposition method was proposed to obtain typical load profiles and detect outliers. The method was tested on a large real dataset to prove the effectiveness and efficiency. Deep-learning-based stacked sparse auto-encoders were applied for load profile compression and feature extraction in \cite{varga2015robust}. Based on the reduced and encoded load profile, a locality sensitive hashing (LSH) method was further proposed to classify the load profiles and obtain the representative load profiles.

Insights into the local and global characteristics of smart meter data are important for finding meaningful typical load profiles. Three new types of features generated by applying conditional filters to meter-resolution-based features integrated with shape signatures, calibration and normalization, and profile errors were proposed in \cite{al2016feature} to cluster daily load curves. The proposed feature extraction method was of low computational complexity, and the features were informative and understandable for describing the electricity usage patterns. To capture local and global shape variations, 10 subspace clustering and projected clustering methods were applied to identify the contact type of consumers in \cite{piao2014subspace}. By focusing on the subspace of load profiles, the clustering process was proven to be more robust to noise. To capture the peak load and major variability in residential consumption behavior, four key time periods (overnight, breakfast, daytime, and evening) were identified in \cite{haben2016analysis}. On this basis, seven attributes were calculated for clustering. The robustness of the proposed clustering was verified using the bootstrap technique.

The variability and uncertainty of smart meter data have also been considered for load profiling.  Four key time periods, which described different peak demand behaviors, coinciding with common intervals of the day were identified in \cite{haben2016analysis}, and then a finite mixture-model-based clustering was used to discover ten distinct behavior groups describing customers based on their demand and variability. The load variation was modeled by a lognormal distribution, and a Gaussian mixture model (GMM)-based load profiling method was proposed in \cite{stephen2014enhanced} to capture dynamic behavior of consumers. A mixture model was also used in \cite{sun2016c} by integrating the C-vine copula method (C-vine copula-based mixture model) for the clustering of residential load profiles. The high-dimensional nonlinear correlations among consumptions of different time periods were modeled using the C-vine copula. This method has an effective performance in large data sets. While in \cite{wang2016clustering}, a Markov model was established based on the separated time periods to describe the electricity consumption behavior dynamics. A clustering technique consisting of fast search and find of density peaks (CFSFDP) integrated into a divide-and-conquer distributed approach was proposed to find typical consumption behaviors. The proposed distributed clustering algorithm had higher computational efficiency. The expectation maximization (EM)-based mixture model clustering method was applied in \cite{labeeuw2013residential} to obtain typical load profiles, and then the variabilities in residential load profiles were modeled by a transition matrix based on a second-order Markov chain and Markov decision processes. The proposed method can be used to generate pseudo smart meter data for retailers and protect the privacy of consumers.

\subsection{Remarks}
Table \ref{Abnormal} provides the correspondence between the key techniques and the surveyed references in smart meter data analytics for load analysis.
\begin{table*}[h]
\caption{Brief Summary of the Literature on Load Analysis}
\label{Abnormal}
\begin{center}
\begin{tabular}{p{3.2cm}<{\centering}|p{5cm}<{\centering}|p{6cm}<{\centering}}
\hline
{\textbf{Load Analysis}} & {\textbf{Key words}} & {\textbf{References}} \\
\hline
\multirow{3}{*}{Bad Data Detection} & Time Series Analysis  & \cite{peppanen2016handling}
\cite{akouemo2017data}
\cite{li2010classification}  \\
\multirow{1}{*}{} & Low Rank Matrix  & \cite{mateos2013load}
\cite{wang2016application}\cite{huang2017false}\\
\multirow{1}{*}{} & Time Window  &  \cite{al2016k} 
\cite{al2016state}
 \cite{araya2017ensemble}
 \cite{liu2016online}\\
\hline
\multirow{2}{*}{Energy Theft Detection} & Supervised Learning & \cite{jokar2016electricity}
\cite{jindal2016decision}
 \cite{depuru2013high} \\
{} & Unsupervised Learning  & \cite{junior2016unsupervised}
\cite{nizar2008power}
\cite{botev2016detecting}
\cite{janetzko2014anomaly}
\\
\hline
\multirow{4}{*}{Load Profiling} & Direct Clustering  & \cite{chicco2012overview} \cite{zhou2013review} \cite{wang2015load} \cite{granell2015impacts}
\cite{benitez2014dynamic} \cite{7947198} \\
\multirow{1}{*}{} & Dimension Reduction & \cite{chicco2012overview} \cite{koivisto2013clustering}
\cite{chelmis2015big}
\cite{varga2015robust}\\
\multirow{1}{*}{} & Local Characteristics & \cite{al2016feature}
\cite{piao2014subspace}
\cite{haben2016analysis}\\
& Variability and Uncertainty & \cite{haben2016analysis}
\cite{wang2016clustering}
\cite{stephen2014enhanced}
 \cite{sun2016c} 
\cite{labeeuw2013residential}\\
\hline
\end{tabular}
\end{center}
\end{table*}

For bad data detection, most of the bad data detection methods are suitable for business/industrial consumers or higher aggregation level load data, which are more regular and have certain patterns. The research on bad data detection on the individual consumer is still limited and not a trivial task because the load profiles of an individual consumer show more variation. In addition, since bad data detection and repairing are the basis of other data analytics application, how much improvement can be made for load forecasting or other applications after bad data detection is also an issue that deserves further investigation. In addition, smart meter data are essentially streaming data. Real-time bad data detection for some real-time applications, such as very-short-term load forecasting, is another concern. Finally, as stated above, bad data may be brought from data collection failure. Short period anomaly usage patterns may also be identified as bad data even through it is "real" data. More related factors such as sudden events need to be considered in this situation. Redundant data are also good sources for "real" but anomaly data identification.

For energy theft detection, with a longer time period of smart meter data, the detection accuracy is probably higher because more data can be used. However, using longer historical smart meter data may also lead to a detection delay, which means that we need to achieve a balance between the detection accuracy and detection delay. Moreover, different private data and simulated data have been tested on different energy theft detection methods in the existed literature. Without the same dataset, the superiority of a certain method cannot be guaranteed. The research on this area will be promoted if some open datasets are provided. Besides, in most cases, one paper proposes one energy theft detection method. Just like ensemble learning for load forecasting, can we propose an ensemble detection framework to combine different individual methods?

For load profiling, the  majority of the clustering methods are used for stored smart meter data. However, the fact is that smart meter data are streaming data. Sometime, we need deal with the massive streaming data in a real-time fashion for specific applications. Thus, distributed clustering and incremental clustering methods can be further studied in the field of load profiling. Indirect load profile methods extract features first and then conduct clustering on the extracted features. Some clustering methods such as deep embedding clustering \cite{xie2016unsupervised} that can implement feature extraction and clustering at the same time, have been proposed outside the area of electrical engineering. It is worth trying to apply these state-of-the-art methods to load profiling. Most load profiling methods are evaluated by clustering-based indices, such as similarity matrix indicator (SMI), Davies-Bouldin indicator (DBI) and Silhouette Index (SIL) \cite{zhang2012new}. More application-oriented matrices such as forecasting accuracy are encouraged to be used to guide the selection of suitable clustering methods.  Finally, how to effectively extract meaningful features before clustering to improve the performance and efficiency of load profiling is another issue that needs to be further addressed.

\section{Load Forecasting}
\label{forecasting}
\begin{figure*}[!t]
  \begin{center}
  \includegraphics[width=7in]{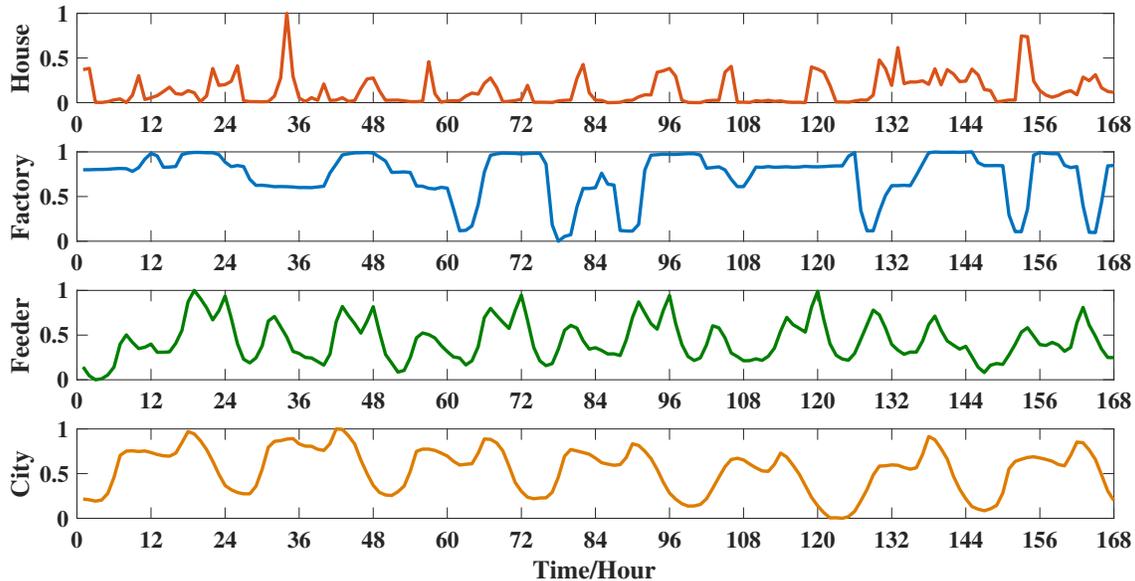}\\
  \caption{Normalized hourly profiles of a week for four types of loads.}\label{LoadLevel}
  \end{center}
\end{figure*}

Load forecasts have been widely used by the electric power industry. Power distribution companies rely on short- and long-term forecasts at the feeder level to support operations and planning processes, while retail electricity providers make pricing, procurement and hedging decisions largely based on the forecasted load of their customers. Fig. \ref{LoadLevel} presents the normalized hourly profiles of a week for four different types of loads, including a house, a factory, a feeder, and a city. The loads of a house, a factory, and a feeder are more volatile than the city-level load. In reality, the higher level the load is measured at, the smoother the load profile typically is. Developing a highly accurate forecast is nontrivial at lower levels. 

Although the majority of the load forecasting literature has been devoted to forecasting at the top (high voltage) level, the information from medium/low voltage levels, such as distribution feeders and even down to the smart meters, offer some opportunities to improve the forecasts. A recent review of load forecasting was conducted in \cite{hong2016probabilistic}, focusing on the transition from point load forecasting to probabilistic load forecasting. In this section, we will review the recent literature for both point and probabilistic load forecasting with the emphasis on the medium/low voltage levels. Within the point load forecasting literature, we divide the review based on whether the smart meter data is used or not.

\subsection{Forecasting without Smart Meter Data}
Compared with the load profiles at the high voltage levels, the load profiles aggregated to a customer group or medium/low voltage level are often more volatile and sensitive to the behaviors of the customers being served. Some of them, such as the load of a residential community, can be very responsive to the weather conditions. Some others, such as the load of a large factor, can be driven by specific work schedules. Although these load profiles differ by the customer composition, these load forecasting problems share some common challenges, such as accounting the influence from the competitive markets, modeling the effects of weather variables, and leveraging the hierarchy. 

In competitive retail markets, the electricity consumption is largely driven by the number of customers. The volatile customer count contributes to the uncertainties in the future load profile. A two-stage long-term retail load forecasting method was proposed in \cite{xie2015long} to take customer attrition into consideration. The first stage was to forecast each customer's load using multiple linear regression with a variable selection method. The second stage was to forecast customer attrition using survival analysis. Thus, the product of the two forecasts provided the final retail load forecast.  
Another issue in the retail market is the consumers' reactions to the various demand response programs. While some consumers may respond to the price signals, others may not. A nonparametric test was applied to detect the demand responsive consumers so that they can be forecasted separately \cite{hoiles2015nonparametric}.  Because the authors did not find publicly available demand data for individual consumers, the experiment was conducted using aggregate load in the Ontario power gird. 

Since the large scale adoption of electrical air conditioning systems in the 1940s, capturing the effects of weather on load has been a major issue in load forecasting. Most load forecasting models in the literature include temperature variables and their variants, such as lags and averages. How many lagged hourly temperatures and moving average temperatures can be included in a regression model? An investigation was conducted in \cite{wang2016electric}. The case study was based on the data from the load forecasting track of GEFCom2012. An important finding is that a regression-based load forecasting model estimated using two to three years of hourly data may include more than a thousand parameters to maximize the forecast accuracy. In addition, each zone may need a different set of lags and moving averages. 

Not many load forecasting papers are devoted to other weather variables. How to include humidity information in load forecasting models was discussed in \cite{xie2016relative}, where the authors discovered that the temperature-humidity index (THI) may not be optimal for load forecasting models. Instead, separating relative humidity, temperature and their higher order terms and interactions in the model, with the corresponding parameters being estimated by the training data, were producing more accurate load forecasts than the THI-based models. A similar investigation was performed for wind speed variables in \cite{xie2017wind}. Comparing with the models that include wind chill index (WCI), the ones with wind speed, temperature, and their variants separated were more accurate. 

The territory of a power company may cover several micro-climate zones. Capturing the local weather information may help improve the load forecast accuracy for each zone. Therefore, proper selection of weather stations would contribute to the final load forecast accuracy. Weather station selection was one of the challenges designed into the load forecasting track of GEFCom2012 \cite{Hong2014Global}. All four winning team adopted the same strategy: first deciding how many stations should be selected, and then figuring out which stations to be selected \cite{charlton2014refined}\cite{lloyd2014gefcom2012}\cite{nedellec2014gefcom2012}\cite{taieb2014gradient}. A different and more accurate method was proposed in \cite{hong2015weather}, which follows a different strategy, determining how many and which stations to be selected at the same time instead of sequentially. The method includes three steps: rating and ranking the individual weather stations, combining weather stations based on a greedy algorithm, and rating and ranking the combined stations. The method is currently being used by many power companies, such as North Carolina Electric Membership Corporation, which was used as one of the case studies in \cite{hong2015weather}.

The pursue of operational excellence and large-scale renewable integration is pushing load forecasting toward the grid edge. Distribution substation load forecasting becomes another emerging topic. One approach is to adopt the forecasting techniques and models with a good performance at higher levels. For instance, a three-stage methodology, which  consists of preprocessing, forecasting, and postprocessing, was taken to forecast loads of three datasets ranging from distribution level to transmission level \cite{hoverstad2015short}. A semi-parametric additive model was proposed in \cite{fan2012short} to forecast the load of Australian National Electricity Market. The same technique was also applied to forecast more than 2200 substation loads of the French distribution network in \cite{goude2014local}. Another load forecasting study on seven substations from the French network was reported in \cite{ding2015next}, where a conventional time series forecasting methodology was used.  The same research group then proposed a neural network model to forecast the load of two French distribution substations, which outperformed a time series model \cite{ding2016neural}.

Another approach to distribution load forecasting is to leverage the connection hierarchy of the power grid. In \cite{sun2016efficient}, The load of a root node of any subtree was forecasted first. The child nodes were then treated separately based on their similarities. The forecast of a "regular" node was proportional to the parent node forecast, while the "irregular" nodes were forecasted individually using neural networks. Another attempt to make use of the hierarchical information for load forecasting was made in \cite{borges2013evaluating}. Two case studies were conducted, one based on New York City and its substations, and the other one based on PJM and its substations. The authors demonstrated the effectiveness of aggregation in improving the higher level load forecast accuracy.

\subsection{Forecasting with Smart Meter Data}
The value that smart meters bring to load forecasting is two-fold. First, smart meters make it possible for the local distribution companies and electricity retailers to better understand and forecast the load of an individual house or building. Second, the high granularity load data provided by smart meters offer great potential for improving the forecast accuracy at aggregate levels.  

Because the electricity consumption behaviors at the household and building levels can be much more random and volatile than those at aggregate levels, the traditional techniques and methods developed for load forecasting at an aggregate level may or may not be well suited. To tackle the problem of smart meter load forecasting, the research community has taken several different approaches, such as evaluating and modifying the existing load forecasting techniques and methodologies, adopting and inventing new ones, and a mixture of them. 

A highly cited study compared seven existing techniques, including linear regression, ANN, SVM and their variants \cite{edwards2012predicting}. The case study was performed based on two datasets: one containing two commercial buildings and the other containing three residential homes. The study demonstrated that these techniques could produce fine forecasts for the two commercial buildings but not the three residential homes. A self-recurrent wavelet neural network (SRWNN) was proposed to forecast an education building in a microgrid setting \cite{chitsaz2015short}. The proposed SRWNN was shown to be more accurate than its ancestor wavelet neural network (WNN) for both building-level load forecasting (e.g., a 694 kW peak education building in British Columbia, Canada) and state- or province-level load forecasting (e.g., British Columbia and California).

Some researchers tried deep learning techniques for the household- and building-level load forecasting. Conditional Restricted Boltzmann Machine (CRBM) and Factored
Conditional Restricted Boltzmann Machine (FCRBM) were assessed in \cite{mocanu2016deep} to estimate energy consumption for a household and three submetering measurements. FCRBM achieves the highest load forecast accuracy compared with ANN, RNN, SVM, and CRBM. Different resolutions ranging from one minute to one week have been tested. A pooling-based deep recurrent neural network (RNN) was proposed in \cite{shi2017deep} to learn spatial information shared between interconnected customers and to address the over-fitting challenges. It outperformed ARIMA, SVR, and classical deep RNN on the Irish CER residential dataset. 

Spartsity is a key character in household level load forecasting. A spatio-temporal forecasting approach was proposed in \cite{tascikaraoglu2016short}, which incorporated a large dataset of many driving factors of the load for all surrounding houses of a target house. The proposed method combined ideas from Compressive Sensing and data decomposition to exploit the low-dimensional structures governing the interactions among the nearby houses. The Pecan Street data was used to evaluate the proposed method. Sparse coding was used to model the usage patterns in \cite{yu2017sparse}. The case study was based on a dataset collected from 5000 households in Chattanooga, TN, where Including the sparse coding features led to 10\% improvements in forecast accuracy. A least absolute shrinkage and selection (LASSO)-based sparse linear method was proposed to forecast individual consumption in \cite{li2017sparse}. The consumer's usage patterns can be extracted from the non-zero coefficients, and it was proven that data from other consumers contribute to the fitted residual. Experiments on real data from Pacific Gas and Electric Company showed that the LASSO-based method has low computational complexity and comparable accuracy.

A commonly used method to reduce noise in smart meter data is to aggregate the individual meters. To keep the salient features from being buried during aggregation, clustering techniques are often used to group similar meters. 
In \cite{chaouch2014clustering}, next-day load forecasting was formulated as a functional time series problem. Clustering was first performed to classify the historical load curves into different groups. The last observed load curve was then assigned to the most similar cluster. Finally, based on the load curves in this cluster, a functional wavelet-kernel (FWK) approach was used to forecast the next-day load curve. The results showed that FWK with clustering outperforms simple FWK. Clustering was also conducted in \cite{hsiao2015household} to obtain the load patterns. Classification from contextual information, including time, temperature, date, and economic indicator to clusters, was then performed. Based on the trained classifier, the daily load can be forecasted with known contextual information. A shape-based clustering method was performed in \cite{teeraratkul2017shape} to capture the time drift characteristic of the individual load, where the cluster number was smaller than those obtained by traditional Euclidean-distance-based clustering methods. The clustering method is quite similar to k-means, while the distance is quantified by dynamic time warping (DTW). Markov models were then constructed to forecast the shape of the next-day load curve. Similar to the clustering method proposed in \cite{teeraratkul2017shape}, a k-shape clustering was proposed in \cite{yang2017k} to forecast building time series data, where the time series shape similarity was used to update the cluster memberships to address the time-drift issue.

The fine-grained smart meter data also introduced new perspectives to the aggregation level load forecasting. A clustering algorithm can be used to group the customers. Each customer group can then be forecasted with different forecasting models. Finally, the aggregated load forecast can be obtained by summing the load forecast of each group. Two datasets including the Irish CER residential dataset and another dataset from New York were used to build the case study in \cite{quilumba2015using}. Both showed that forecast errors can be reduced by effectively grouping different customers based on their energy consumption behaviors. A similar finding was presented in \cite{wijaya2015cluster} where the Irish CER residential dataset was used in the case study. The results showed that cluster-based forecasting can improve the forecasting accuracy and that the performance depends on the number of clusters and the size of the consumer. 

The relationship between group size and forecast accuracy based on Seasonal-Naïve and Holt-Winters algorithms was investigated in \cite{da2014impact}. The results showed that forecasting accuracy increases as group size increases, even for small groups. A simple empirical scaling law is proposed in \cite{sevlianscaling} to describe how the accuracy changes as different aggregation levels. The derivation of the scaling law is based on Mean Absolute Percentage Error (MAPE). Case studies on the data from Pacific Gas and Electric Company show that MAPE decreases quickly with the increase of the number of consumers when the number of consumers is less than 100,000. When the number of consumers is more than 100,000, the MAPE has a little decrease.  

Forecast combination is a well-known approach to accuracy improvement. A residential load forecasting case study showed that the ensembles outperformed all the individual forecasts from traditional load forecasting models \cite{stephen2015incorporating}. By varying the number of clusters, different forecasts can be obtained. A novel ensemble forecasting framework was proposed in \cite{wangensemble} to optimally combine these forecasts to further improve the forecasting accuracy.

Traditional error measures such as MAPE cannot reasonably quantify the performance of individual load forecasting due to the violation and time-shifting characteristics. For example, MAPE can easily be influenced by outliers. A resistant MAPE (r-MAPE) based on the calculation of the Huber M-estimator was proposed in \cite{moreno2013using} to overcome this situation. The mean arctangent absolute percentage error (MAAPE) was proposed in \cite{kim2016new} to consider the intermittent nature of individual load profiles. MAAPE, a variation of MAPE, is a slope as an angle, the tangent of which is equal to the ratio between the absolute error and real value, i.e., absolute percentage error (APE). An error measure designed for household level load forecasts was proposed in \cite{haben2014new} to address the time-shifting characteristic of household level loads. In addition to these error measures, some modifications of MAPE and mean absolute error (MAE) have been used in other case studies \cite{yu2017sparse}\cite{li2017sparse}.

\subsection{Probabilistic Forecasting}
\begin{figure}[!t]
  \begin{center}
  \includegraphics[width=3.6in]{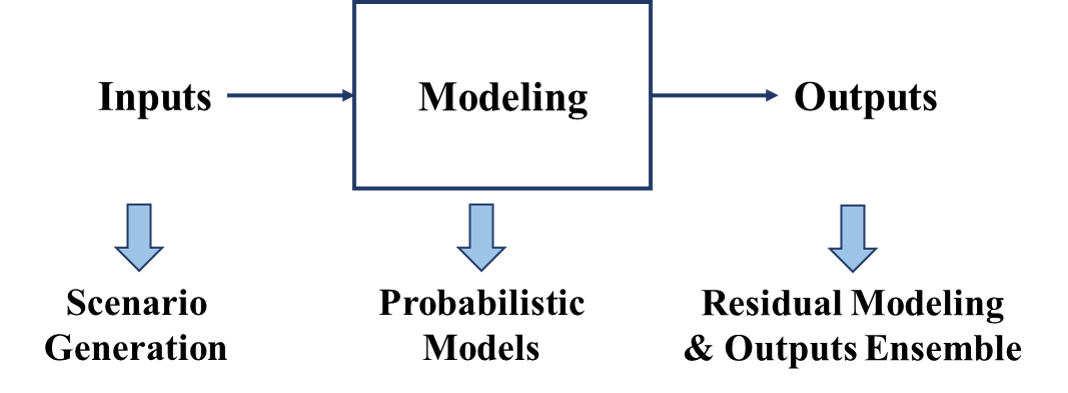}\\
  \caption{From point forecasting to probabilistic forecasting.}\label{Probabilistic}
  \end{center}
\end{figure}

A probabilistic forecast provides more information about future uncertainties that what a point forecast does. As shown in Fig. \ref{Probabilistic}, a typical point forecasting process contains three parts: data inputs, modeling, and data outputs (forecasts). As summarized in \cite{hong2016probabilistic}, there are three ways to modify the workflow to generate probabilistic forecasts: 1) generating multiple input scenarios to feed to a point forecasting model; 2) applying probabilistic forecasting models, such as quantile regression; and 3) augmenting point outputs to probabilistic outputs by imposing simulated or modeled residuals or making ensembles of point forecasts. 

On the input side, scenario generation is an effective way to capture the uncertainties from the driving factors of electricity demand. Various temperature scenario generation methods have been proposed in the literature, such as direct usage of the previous years of hourly temperatures with the dates fixed \cite{hong2014long}, shifting the historical temperatures by a few days to create additional scenarios \cite{PJM}, and bootstrapping the historical temperatures \cite{hyndman2010density}. A comparison of these three methods based on pinball loss function was presented in  \cite{xie2016temperature}. The results showed that the shifted-date method dominated the other two when the number of dates being shifted is within a range. An empirical formula was also proposed to select parameters for the temperature scenario generation methods. The idea of generating temperature scenarios was also applied in \cite{wangembedding}. An embedding based quantile regression neural network was used as the regression mode instead of MLR model, where the embedding layer can model the effect of calendar variables. In this way, the uncertainties of both future temperature and the relationship between temperature and load can be comprehensively considered. The scenario generation method was also used to develop a probabilistic view of power distribution system reliability indices \cite{PEM2}. 

On the output side, one can convert point forecasts to probabilistic ones via residual simulation or forecast combination. Several residual simulation methods were evaluated in \cite{7163624}. The results showed that the residuals do not always follow a normal distribution, though group analysis increases the passing rate of normality tests. Adding simulated residuals under the normality assumption improves probabilistic forecasts from deficient models, while the improvement is diminishing as the underlying model improves. The idea of combining point load forecasts to generate probabilistic load forecasts was first proposed in \cite{liu2015probabilistic}. The quantile regression averaging (QRA) method was applied to eight sister load forecasts, a set of point forecasts generated from homogeneous models developed in \cite{wang2016electric}. A constrained QRA (CQRA) was proposed in \cite{combining} to combine a series of quantiles obtained from individual quantile regression models.

Both approaches mentioned above rely on point forecasting models. It is still an unsolved question whether a more accurate point forecasting model can lead to a more skilled probabilistic forecast within this framework. An attempt was made in \cite{7922570} to answer this question. The finding is that when the two underlying models are significantly different w.r.t. the point forecast accuracy, a more accurate point forecasting model would lead to a more skilled probabilistic forecast. 

Various probabilistic forecasting models have been proposed by statisticians and computer scientists, such as quantile regression, Gaussian process regression, and density estimation. These off-the-shelf models can be directly applied to generate probabilistic load forecasts \cite{hong2016probabilistic}. In GEFCom2014, a winning team developed a quantile generalized additive model (quantGAM), which is a hybrid of quantile regression and generalized additive models \cite{gaillard2016additive}. Probabilistic load forecasting has also been conducted on individual load profiles. Combining the gradient boosting method and quantile regression, a boosting additive quantile regression method was proposed in \cite{taieb2016forecasting} to quantify the uncertainty and generate probabilistic forecasts. Apart from the quantile regression model, kernel density estimation methods were tested in \cite{arora2016forecasting}. The density of electricity data was modeled using different implementations of conditional kernel density (CKD) estimators to accommodate the seasonality in consumption. A decay parameter was used in the density estimation model for recent effects. The selection of kernel bandwidths and the presence of boundary effects are two main challenges with the implementation of CKD that were also investigated.

\subsection{Remarks}
Table \ref{Forecasting} provides the correspondence between the key techniques and the surveyed references in smart meter data analytics for load forecasting.

\begin{table*}[h]
\caption{Brief Summary of the Literature on Load Forecasting}
\label{Forecasting}
\begin{center}
\begin{tabular}{p{3.2cm}<{\centering}|p{5cm}<{\centering}|p{6cm}<{\centering}}
\hline
{\textbf{Load Forecasting}} & {\textbf{Key words}} & {\textbf{References}} \\
\hline
\multirow{4}{*}{Without Individual Meters} & Consumer Attrition / Demand Response& 
\cite{xie2015long}\cite{hoiles2015nonparametric}
\\
& Weather Modeling \& Selection & \cite{wang2016electric}
\cite{xie2016relative}
\cite{xie2017wind}
\cite{Hong2014Global}
\cite{charlton2014refined}
\cite{lloyd2014gefcom2012}
\cite{nedellec2014gefcom2012}
\cite{taieb2014gradient}
\cite{hong2015weather}
\\
& Traditional High Accurate Model & \cite{hoverstad2015short}
\cite{fan2012short}
\cite{goude2014local}
\cite{ding2015next}
\cite{ding2016neural}
\\
& Hierarchical Forecasting & \cite{sun2016efficient}
\cite{borges2013evaluating}
 \\
\hline
\multirow{5}{*}{With Individual Meters} & Traditional Methods& \cite{edwards2012predicting}
\cite{chitsaz2015short}\\
& Sparse Coding / Deep Learning & 
\cite{tascikaraoglu2016short}
\cite{yu2017sparse}
\cite{li2017sparse}
\cite{mocanu2016deep}
 \cite{shi2017deep} \\
& Clustering & \cite{chaouch2014clustering}
\cite{hsiao2015household} 
\cite{teeraratkul2017shape} \cite{yang2017k} \\
& Aggregation Load & \cite{quilumba2015using}
\cite{wijaya2015cluster}\cite{zhang2015short}
\cite{stephen2015incorporating}
\cite{da2014impact}
\cite{humeau2013electricity}\cite{wangensemble} \\
& Evaluation Criteria & \cite{moreno2013using}
\cite{haben2014new} \cite{yu2017sparse} \cite{li2017sparse} \cite{taieb2016forecasting} \cite{kim2016new} \\
\hline
\multirow{3}{*}{Probabilistic Forecasting} & Scenario Generation & \cite{hong2014long}
\cite{PJM}
\cite{hyndman2010density}
\cite{xie2016temperature}
\cite{wangembedding}\cite{PEM2}
\\
& Residual Modeling \& Output Ensemble & \cite{7163624}
\cite{liu2015probabilistic}
\cite{wang2016electric}
\cite{combining}
\\
& Probabilistic Forecasting Models & 
\cite{hong2016probabilistic}
\cite{gaillard2016additive}
\cite{taieb2016forecasting}
\cite{arora2016forecasting}
\\
\hline
\end{tabular}
\end{center}
\end{table*}
Forecasting the loads at aggregate levels is a relatively mature area. Nevertheless, there are some nuances in the smart grid era due to the increasing need of highly accurate load forecasts. One is on the evaluation methods. Many forecasts are being evaluated using widely used error measures such as MAPE, which does not consider the consequences of over- or under-forecasts. In reality, the cost to the sign and maganitude of errors may differ significantly. Therefore, the following research question rises: how can the costs of forecast errors be integrated into the forecasting processes? Some research in this area would be helpful to bridge the gap between forecasting and decision making. The second one is load transfer detection, which is a rarely touched area in the literature. Distribution operators may transfer the load from one circuit to another permanently, seasonally, or on an ad hoc basis, in response to maintenance needs or reliability reasons. These load transfers are often poorly documented. Without smart meter information, it is difficult to physically trace the load blocks being transferred. Therefore, a data-driven approach is necessary in these situations. The third one is hierarchical forecasting,  specifically, how to fully utilize zonal, regional, or meter load and local weather data to improve the load forecast accuracy. In addition, it is worth studying how to reconcile the forecasts from different levels for the applications of aggregators, system operators, and planners. The fourth one is on the emerging factors that affect electricity demand. The consumer behaviors are being changed by many modern technologies, such as rooftop solar panels, large batteries, and smart home devices. It is important to leverage the emerging data sources, such as technology adoption, social media, and various marketing surveys.

To comprehensively capture the uncertainties in the future, researchers and practitioners recently started to investigate in probabilistic load forecasting. Several areas within probabilistic load forecasting would need some further attention. First, distributed energy resources and energy storage options often disrupt the traditional load profiles. Some research is needed to generate probabilistic net load forecasts for the system with high penetration of renewable energy and large scale storage. Secondly, forecast combination is widely regarded in the point forecasting literature as an effective way to enhance the forecast accuracy. There is a primary attempt in \cite{combining} to combine quantile forecasts. Further investigations can be conducted on combining other forms probabilistic forecasts, such as density forecasts and interval forecasts. Finally, the literature of probabilistic load forecasting for smart meters is still quite limited. Since the meter-level loads are more volatile than the aggregate loads, probabilistic forecasting has a natural application in this area. 

\section{Load Management}
\label{demandresponse}
How smart meter data contribute to the implementation of load management is summarized  from three aspects in this section: the first one is to have a better understanding of sociodemographic information of consumers to provide better and personalized service. The second one is to target the potential consumers for demand response program marketing. The third one is the issue related to demand response program implementation including price design for price-based demand response and baseline estimation for incentive-based demand response.

\subsection{Consumer Characterization}
The electricity consumption behaviors of the consumers are closely related to their sociodemographic status. Bridging the load profiles to sociodemographic status is an important approach to classify the consumers and realize personalized services. A naive problem is to detect the consumer types according to the load profiles. The other two issues are identifying sociodemographic information from load profiles and predicting the load shapes using the sociodemographic information. 

Identifying the type of consumers can be realized by simple classification. The temporal load profiles were first transformed into the frequency domain in \cite{zhong2015hierarchical} using fast Fourier transformation (FFT). Then the coefficients of different frequencies were used as the inputs of classification and regression tree (CART) to place consumers in different categories. FFT decomposes smart meter data based on a certain sine function and cosine function. Another transformation technique, sparse coding, has no assumption on the base signal but learns them automatically. Non-negative sparse coding was applied to extract the partial usage patterns from original load profiles in \cite{wang2016sparse}.  Based on the partial usage patterns, linear SVM was implemented to classify the consumers into residents and small and medium-sized enterprises (SME). The classification accuracy is considerably higher than discrete wavelet transform (DWT) and PCA.

There are still consumers without smart meter installations. External data, such as the sociodemographic status of consumers, are applied to estimate their load profiles. Clustering was first implemented to classify consumers into different energy behavior groups, and then energy behavior correlation rate (EBCR) and indicator dominance index (IGD) were defined and calculated to identify the indicators higher than a threshold \cite{tong2016cross}. Finally, the relationship between different energy behavior groups and their sociodemographic status was mapped. Spectral clustering was applied to generate typical load profiles, which were then used as the inputs of predictors such as random forests (RF) and stochastic boosting (SB) in \cite{7343745}. The results showed that with commercial and cartographic data, the load profiles of consumers can be accurately predicted. Stepwise selection was applied to investigate the factors that have a great influence on residential electricity consumption in \cite{kavousian2013determinants}.  The location, floor area, the age of consumers, and the number of appliances are main factors, while the income level and home ownership have little relationship with consumption. A multiple linear regression model was used to bridge the total electricity consumption, maximum demand, load factor, and ToU to dwelling and occupant socioeconomic variables in \cite{mcloughlin2012characterising}. The factors that have a great impact on total consumption, maximum load, load factor, and ToU were identified. The influence of socioeconomic status of consumers' electricity consumption patterns was evaluated in \cite{han2014impact}. Random forest (RF) regression was proposed to combine socioeconomic status and environmental factors to predict the consumption patterns.

More works focus on how to mine the sociodemographic information of consumers from the massive smart meter data. One approach is based on a clustering algorithm. DPMM was applied in \cite{granell2015clustering} for household and business premise load profiling where the number of clusters was not required to predetermined. The clustering results obtained by the DPMM algorithm have a clear corresponding relation with the metadata of dwellings, such as the nationality, household size, and type of dwelling. Based on the clustering results, multinomial logistic regression was applied to the clusters and dwelling and appliance characteristics in \cite{mcloughlin2015clustering}. Each cluster was analyzed according to the coefficients of the regression model. 
Feature extraction and selection have also been applied as the attributes of the classifier. A feature set including the average consumption over a certain period, the ratios of two consumptions in different periods, and the temporal properties was established in \cite{beckel2014revealing}. Then, classification or regression was implemented to predict the sociodemographic status according to these features. Results showed that the proposed feature extraction method outperform biased random guess. 
More than 88 features from consumption, ratios, statistics, and temporal characteristics were extracted, and then correlation, KS-test, and ${{\eta }^{2}}$-based feature selection methods were conducted in \cite{hopf2016feature}. The so-called extend CLASS classification framework was used to forecast the deduced properties of private dwellings. A supervised classification algorithm called dependent-independent data classification (DID-Class) was proposed to address the challenges of dependencies among multiple classification-relevant variables in \cite{sodenkamp2016supervised}. The characteristics of dwellings were recognized based on this method, and comparisons with SVM and traditional CLASS proposed in \cite{beckel2014revealing} were conducted. The accuracy of DID-Class with SVM and CLASS is slightly higher than those of SVM and CLASS. To capture the intra-day and inter-day electricity consumption behavior of the consumers, a two-dimensional convolutional neural network (CNN) was used in \cite{8291011} to make a bridge between the smart meter data and sociodemographic information of the consumers. The deep learning method can extract the features automatically and outperforms traditional methods.

\subsection{Demand Response Program Marketing}
Demand response program marketing is to target the consumers who have a large potential to be involved in demand response programs. On one hand, 15 minute or half-hour smart meter data cannot provide detail information on the operation status of the appliance; on the other hand, the altitude of consumers towards demand response is hard to model. Thus, the demand response potential cannot be evaluated directly. In this subsection, the potential of demand response can be indirectly evaluated by analyzing the variability, sensitivity to temperature, and so forth.

Variability is a key index for evaluating the potential of demand response. A hidden Markov model (HMM)-based spectral clustering was proposed in \cite{albert2013smart} to describe the magnitude, duration, and variability of the electricity consumption and further estimate the occupancy states of consumers. The information on the variability, occupancy states, and inter-temporal consumption dynamics can help retailers or aggregators target suitable consumers at different time scales. Both adaptive k-means and hierarchical clustering were used to obtain the typical load shapes of all the consumers within a certain error threshold in \cite{kwac2014household}. The entropy of each consumer was then calculated according to the distribution of daily load profiles over a year, and the typical shapes of load profiles were analyzed. The consumers with lower entropy have relatively similar consumption patterns on different days and can be viewed as a greater potential for demand response because their load profiles are more predictable. Similarly, the entropy was calculated in \cite{wang2016clustering} based on the state transition matrix. It was stated that the consumers with high entropy are suitable for price-based demand response for their flexibility to adjust their load profile according to the change in price, whereas the consumers with low entropy are suitable for incentive-based demand response for their predictability to follow the control commands. 

Estimation of electricity reduction is another approach for demand response potential. A mixture model clustering was conducted on a survey dataset and smart meter data in \cite{labeeuw2013residential} to evaluate the potential for active demand reduction with wet appliances. The results showed that both the electricity demand of wet appliances and the attitudes toward demand response have a great influence on the potential for load shifting. Based on the GMM model of the electricity consumption of consumers and the estimated baseline, two indices, i.e., the possibility of electricity reduction greater than or equal to a certain value and the least amount of electricity reduction with a certain possibility, were calculated in \cite{bai2016real}. These two indices can help demand response implementers have a probabilistic understanding of how much electricity can be reduced. A two-stage demand response management strategy was proposed in \cite{jindal2016data}, where SVM was first used to detect the devices and users with excess load consumption and then a load balancing algorithm was performed to balance the overall load.

Since appliances such as heating, ventilation and air conditioning (HVAC) have great potential for demand response, the sensitivity of electricity consumption to outdoor air temperature is an effective evaluation criterion. Linear regression was applied to smart meter data and temperature data to calculate this sensitivity, and the maximum likelihood approach was used to estimate the changing point in \cite{dyson2014using}. Based on that, the demand response potentials at different hours were estimated. Apart from the simple regression, an HMM-based thermal regime was proposed to separate the original load profile into the thermal profile (temperature-sensitive) and base profile (non-temperature-sensitive) in \cite{albert2015thermal}.  The demand response potential can be calculated for different situations, and the proposed method can achieve much more savings than random selection. A thermal demand response ranking method was proposed in \cite{albert2016finding} for demand response targeting, where the demand response potential was evaluated from two aspects: temperature sensitivity and occupancy. Both linear regression and breakpoint detection were used to model the thermal regimes; the true linear response rate was used to detect the occupancy.

\subsection{Demand Response Implementation}
Demand response can be roughly divided into price-based demand response and incentive-based demand response. Price design is an important business model to attract consumers and maximize profit in price-based demand response programs; baseline estimation is the basis of quantifying the performance of consumers in incentive-based demand response programs. The applications of smart meter data analytics in price design and baseline estimation are summarized in this subsection. 

For tariff design, an improved weighted fuzzy average (WFA) K-means was first proposed to obtain typical load profiles in \cite{mahmoudi2010three}. An optimization model was then formulated with a designed profit function, where the acceptance of consumers over price was modeled by a piecewise function. The similar price determination strategy was also presented in \cite{joseph2018real}. Conditional value at risk (CVaR) for the risk model was further considered in \cite{mahmoudi2010annual} such that the original optimization model becomes a stochastic one. 
Different types of clustering algorithms were applied to extract load profiles with a performance index granularity guided in \cite{crow2014clustering}. The results showed that different clusterings with different numbers of clusters and algorithms lead to different costs. GMM clustering was implemented on both energy prices and load profiles in \cite{li2016novel}. Then, ToU tariff was developed using different combinations of the classifications of time periods. The impact of the designed price on demand response was finally quantified. 

For baseline estimation, 
five naive baseline methods, HighXofY, MidXofY, LowXofY, exponential moving average, and regression baselines, were introduced in \cite{wijaya2014bias}. Different demand response scenarios were modeled and considered. The results showed that bias rather than accuracy is the main factor for deciding which baseline provides the largest profits. 
To describe the uncertainty within the consumption behaviors of consumers, Gaussian-process-based probabilistic baseline estimation was proposed in \cite{weng2015probabilistic}. In addition, how the aggregation level influences the relative estimation error was also investigated. K-means clustering of the load profiles in non-event days was first applied in \cite{zhang2016cluster}, and a decision tree was used to predict the electricity consumption level according to demographics data, including household characteristics and electrical appliances. Thus, a new consumer can be directly classified into a certain group before joining the demand response program and then simple averaging and piecewise linear regression were used to estimate to baseline load in different weather conditions. Selecting a control group for baseline estimation was formulated as an optimization problem in \cite{hatton2016statistical}. The objective was to minimize the difference between the load profiles of the control group and demand response group when there is no demand response event. The problem was transformed into a constrained regression problem. 

\subsection{Remarks}
Table \ref{Response} provides the correspondence between the key techniques and the surveyed references in smart meter data analytics for load management.
\begin{table*}[h]
\caption{Brief Summary of the Literature on Load Management}
\label{Response}
\begin{center}
\begin{tabular}{p{3.2cm}<{\centering}|p{5cm}<{\centering}|p{6cm}<{\centering}}
\hline
{\textbf{Load Management}} & {\textbf{Key words}} & {\textbf{References}} \\
\hline
\multirow{3}{*}{Consumer  Characterization} & Consumer Type &  \cite{wang2016sparse}
\cite{zhong2015hierarchical}  \\
& Load Profile Prediction & \cite{tong2016cross}
\cite{7343745}
\cite{kavousian2013determinants}
\cite{mcloughlin2012characterising}
\cite{han2014impact} \\
& Socio-demographic Status Prediction & \cite{granell2015clustering}
\cite{mcloughlin2015clustering}
\cite{beckel2014revealing}\cite{8291011}
\cite{hopf2016feature}
\cite{sodenkamp2016supervised} \\
\hline
\multirow{2}{*}{Demand Response} & Variability  & \cite{albert2013smart}
\cite{kwac2014household}
\cite{wang2016clustering} \\
\multirow{2}{*}{Program Marketing}& Electricity Reduction & \cite{labeeuw2013residential}
\cite{bai2016real}\cite{jindal2016data} \\
& Temperature Sensitivity & \cite{dyson2014using}
\cite{albert2015thermal}\cite{albert2016finding} \\
\hline
\multirow{1}{*}{Demand Response} & Tariff Design &  \cite{mahmoudi2010three}\cite{joseph2018real}
\cite{mahmoudi2010annual}
\cite{crow2014clustering}
\cite{li2016novel}  \\
\multirow{1}{*}{Implementation}& Baseline Estimation & 
\cite{wijaya2014bias}
\cite{weng2015probabilistic}
\cite{zhang2016cluster}
\cite{hatton2016statistical} \\
\hline
\end{tabular}
\end{center}
\end{table*}

For consumer characterization, it is essentially a high dimensional and nonlinear classification problem. There are at least two ways to improve the performance of consumer characterization: 1) conducting feature extraction or selection; 2) developing classification models. In most existing literature, the features for consumer characterization are manually extracted. A data-driven feature extraction method might be an effective way to further improve the performance. The classification is mainly implemented by the shallow learning models such as ANN and SVM. We can try different deep learning networks to tackle the high nonlinearity. We also find that the current works are mainly based on the Irish dataset \cite{Irish}. Low Carbon London dataset may be another good choice. More open datasets are needed to enrich the research in this area. 

For demand response program marketing, evaluating the potential for load shifting or reduction is an effective way to target suitable consumers for different demand response programs. Smart meter data with a frequency of 30 minutes or lower cannot reveal the operation states of the individual appliance; thus, several indirect indices, including entropy, sensitivity to temperature and price, are used. More indices can be further proposed to provide a comprehensive understanding of the electricity consumption behavior of consumers. Since most papers target potential consumers for demand response according to the indirect indices, a critical question is why and how these indices can reflect the demand response potential without experimental evidence? More real-world experimental results are welcomed for the research.

For demand response implementation, all the price designs surveyed above are implemented with a known acceptance function against price. However, the acceptance function or utility function is hard to estimate. How to obtain the function has not been introduced in existing literature. If the used acceptance function or utility function is different from the real one, the obtained results will deviate from the optimal results. Sensitivity analysis of the acceptance function or utility function assumption can be further conducted. Except for traditional tariff design, some innovative prices can be studied, such as different tariff packages based on fine-grained smart meter data. For baseline estimation, in addition to deterministic estimation, probabilistic estimation methods can present more future uncertainties. Another issue is how to effectively incorporate the deterministic or probabilistic baseline estimation results into demand response scheduling problem.

\section{Miscellanies}
\label{miscellanies}
In addition to the three main applications summarized above, the works on smart meter data analytics also cover some other applications, including power network connection verification, outage management, data compression, data privacy, and so forth. Since only several trials have been conducted in these areas and the works in the literature are not so rich, the works are summarized in this miscellanies section.

\subsection{Connection Verification}
The distribution connection information can help utilities and DSO make the optimal decision regarding the operation of the distribution system. Unfortunately, the entire topology of the system may not be available especially at low voltage levels. Several works have been conducted to identify the connections of different demand nodes using smart meter data. 

Correlation analysis of the hourly voltage and power consumption data from smart meters were used to correct connectivity errors in \cite{luan2015smart}. The analysis assumed that the voltage magnitude decreases downstream along the feeder. However, the assumption might be incorrect when there is a large amount of distributed renewable energy integration. In addition to consumption data, both the voltage and current data were used in \cite{peppanen2016distribution} to estimate the topology of the distribution system secondary circuit and the impedance of each branch. This estimation was conducted in a greedy fashion rather than an exhaustive search to enhance computational efficiency. The topology identification problem was formulated as an optimization problem minimizing the mutual-information-based Kullback-Leibler (KL) divergence between each two voltage time series in \cite{weng2016distributed}. The effectiveness of mutual information was discussed from the perspective of conditional probability. Similarly, based on the assumption that the correlation between interconnected neighboring buses is higher than that between non-neighbor buses, the topology identification problem was formulated as a probabilistic graph model and a Lasso-based sparse estimation problem in \cite{liao2016urban}. How to choose the regularization parameter for Lasso regression was also discussed.

The electricity consumption data at different levels were analyzed by PCA in \cite{pappu2017identifying} for both phase and topology identification where the errors caused by technical loss, smart metering, and clock synchronization were formulated as Gaussian distributions.
Rather than using all smart meter data, a phase identification problem with incomplete data was proposed in \cite{xu2016phase} to address the challenge of bad data or null data. The high-frequency load was first obtained by a Fourier transform, and then the variations in high-frequency load between two adjacent time intervals were extracted as the inputs of saliency analysis for phase identification. A sensitivity analysis on smart meter penetration ratios was performed and showed that over 95\% accuracy can be achieved with only 10\% smart meters.

\subsection{Outage Management}
A power outage is defined as an electricity supply failure, which may be caused by short circuits, station failure, and distribution line damage \cite{gungor2013survey}. Outage management is viewed as one of the highest priorities of smart meter data analytics behind billing. It includes outage notification (or last gasp), outage location and restoration verification.

How the outage management applications work, the data requirements, and the system integration considerations were introduced in \cite{tram2008technical}. The outage area was identified using a two-stage strategy in \cite{he2016smart}. In the first stage, the physical distribution network was simplified using topology analysis; in the second stage, the outage area was identified using smart meter information, where the impacts of communication were also considered. A smart meter data-based outage location prediction method was proposed in \cite{kuroda2014approach} to rapidly detect and recover the power outages. The challenges of smart meter data utilization and required functions were analyzed. Additionally, as a way to identify the faulted section on a feeder or lateral, a new multiple-hypothesis method was proposed in \cite{jiang2016outage}, where the outage reports from smart meters were used as the input of the proposed multiple-hypothesis method. The problem was formulated as an optimization model to maximize the number of smart meter notifications. A novel hierarchical framework was established in \cite{moghaddass2017hierarchical} for outage detection using smart meter event data rather than consumption data. It can address the challenges of missing data, multivariate count data, and variable selection. How to use data analytics method to model the outages and reliability indices from weather data was discussed in \cite{PEM2}. Apart from the data analytics method for outage management, more works on smart meter data-based outage managements have been adopted to the corresponding communication architectures \cite{zheng2013smart}\cite{andrysiak2017anomaly}.

\subsection{Data Compression}
Massive smart meter data present more challenges with respect to data communication and storage. Compressing smart meter data to a very small size and without (large) loss can ease the communication and storage burden. Data compression can be divided into lossy compression and lossless compression. Different compression methods for electric signal waveforms in smart grids are summarized in \cite{tcheou2014compression}.

Some papers exist that specifically discuss the smart meter data compression problem.
Note that the changes in electricity consumption in adjunct time periods are much smaller than the actual consumption, particularly for very high-frequency data. Thus, combining normalization, variable-length coding, and entropy coding, and the differential coding method was proposed in \cite{unterweger2015resumable} for the lossless compression of smart meter data. While different lossless compression methods, including IEC 62056-21, A-XDR, differential exponential Golomb and arithmetic (DEGA) coding, and Lempel Ziv Markov chain algorithm (LZMA) coding, were compared on REDD and SAG datasets in \cite{unterweger2015effect}. The performances on the data with different granularities were investigated. The results showed that these lossless compression methods have better performance on higher granularity data.

For low granularity (such as 15 minutes) smart meter data,  symbolic aggregate approximation (SAX), a classic time series data compression method, was used in \cite{notaristefano2013data} and \cite{wang2016clustering} to reduce the dimensionality of load profiles before clustering. The distribution of load profiles was first fitted by generalized extreme value in \cite{tong2016smart}. A feature-based load data compression method (FLDC) was proposed through defining the base state and stimulus state of the load profile and detecting the change in load status. Comparisons with the piecewise aggregate approximation (PAA), SAX, and discrete wavelet transform (DWT) were conducted. Non-negative sparse coding was applied to transform original load profiles into a higher dimensional space in \cite{wang2016sparse} to identify the partial usage patterns and compress the load in a sparse way.

\subsection{Data Privacy}
One of the main oppositions and concerns for the installation of smart meters is the privacy issue. The sociodemographic information can be inferred from the fine-grained smart meter data, as introduced in Section \ref{demandresponse}. There are several works in the literature that discuss how to preserve the privacy of consumers. 

A study on the distributed aggregation architecture for additive smart meter data was conducted in \cite{rottondi2013distributed}. A secure communication protocol was designed for the gateways placed at the consumers’ premises to prevent revealing individual data information. The proposed communication protocol can be implemented in both centralized and distributed manners.
A framework for the trade-off between privacy and utility requirement of consumers was presented in \cite{sankar2013smart} based on a hidden Markov model. The utility requirement was evaluated by the distortion between the original and the perturbed data, while the privacy was evaluated by the mutual information between the two data sequences. Then, a utility-privacy trade-off region was defined from the perspective of information theory. This trade-off was also investigated in \cite{savi2015evaluation}, where the attack success probability was defined as an objective function to be minimized and $\varepsilon$-privacy was formulated. The aggregation of individual smart meter data and the introduction of colored noise were used to reduce the success probability.

Edge detection is one main approach for NILM to identify the status of appliances.  How the data granularity of smart meter data influences the edge detection performance was studied in \cite{eibl2015influence}. The results showed that when the data collection frequency is lower than half the on-time of the appliance, the detection rate dramatically decreases. The privacy was evaluated by the F-score of NILM. The privacy preservation problem was formulated as an optimization problem in \cite{7956260}, where the objective was to minimize the sum of the expected cost, disutility of consumers caused by the late use of appliances, and information leakage. Eight privacy-enhanced scheduling strategies considering on-site battery, renewable energy resources, and appliance load moderation were comprehensively compared.

\section{Open Research Issues}
\label{openissue}
Although smart meter data analytics has received extensive attention and rich literature studies related to this area have been published, developments in computer science and the energy system itself will certainly lead to new problems or opportunities. In this section, several works on smart meter data analytics in the future smart grid are highlighted.

\subsection{Big Data Issues}
\begin{figure*}[!t]
  \begin{center}
  \includegraphics[width=6.5in]{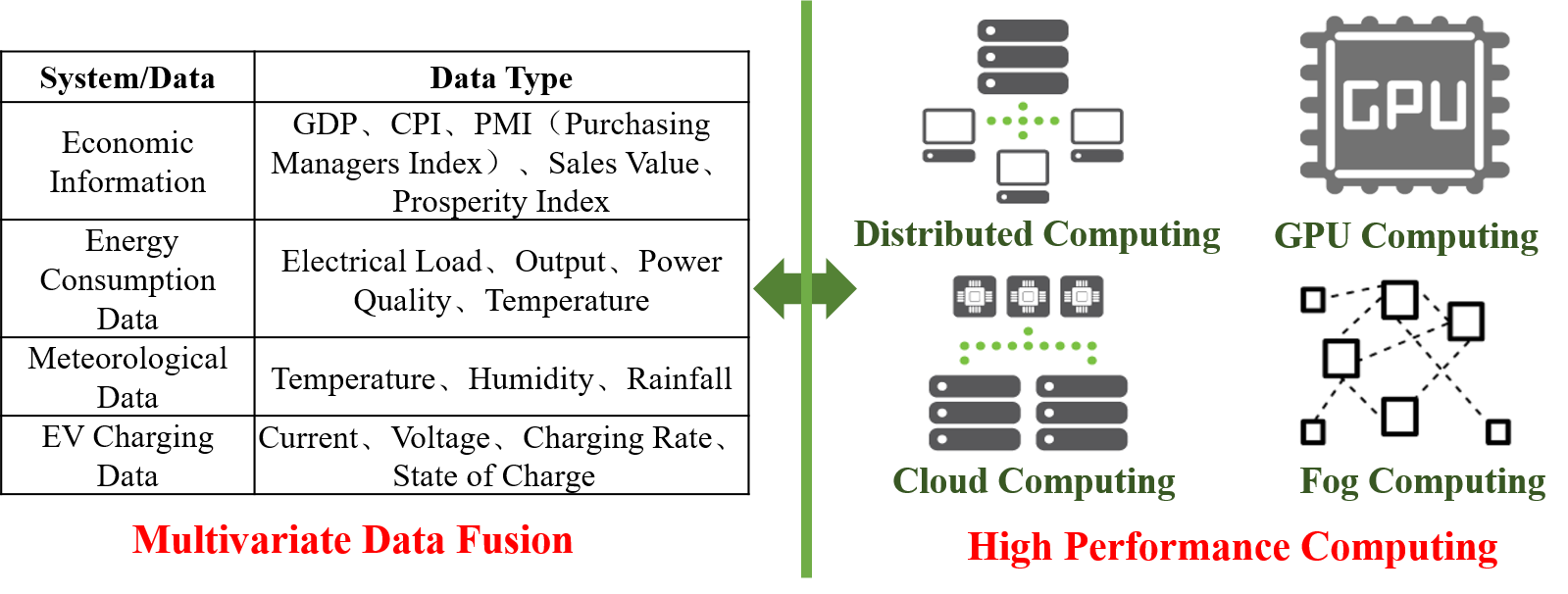}\\
  \caption{Big data issues with smart meter data analytics.}\label{BigData}
  \end{center}
\end{figure*}

Substantial works in the literature have conducted smart meter data analytics. Two special sections about big data analytics for smart grid modernization were hosted in IEEE Transactions on Smart Grid in 2016 \cite{hong2016guest} and IEEE Power and Energy Magazine in 2018, respectively \cite{PEM1}.  However, the size of the dataset analyzed can hardly be called big data. 
How to efficiently integrate more multivariate data with a larger size to discover more knowledge is an emerging issue. As shown in Fig. \ref{BigData}, big data issues with smart meter data analytics include at least two aspects: the first is multivariate data fusion, such as economic information, meteorological data, and EV charging data apart from energy consumption data; the second is high-performance computing, such as distributed computing, GPU computing, cloud computing, and fog computing. 

\subsubsection{Multivariate Data Fusion}
The fusion of various data is one of the basic characteristics of big data \cite{song2017next}. Current studies mainly only focus on the smart meter data itself or even electricity consumption data. Very few papers consider weather data, survey data from consumers, or some other data. Integrating more external data, such as crowd-sourcing data from the Internet, weather data, voltage and current data, and even voice data from service systems may reveal more information. The multivariate data fusion needs to deal with structured data with different granularities and unstructured data. We would like to emphasis that big data is a change of concept. More data-driven methods will be proposed to solve practical problems that may traditionally be solved by model-based methods. For example, with redundant smart meter data, the power flow of the distribution system can be approximated through hyperplane fitting methods such as ANN and SVM. In addition, how to visualize high dimensional and multivariate data to highlight the crucial components and discover the hidden patterns or correlations among these data is a very seldom touched area \cite{PEM3}. 

\subsubsection{High-Performance Computation}
In addition, a majority of smart meter data analytics methods that are applicable to small data sets may not be appropriate for large data sets. Highly efficient algorithms and tools such as distributed and parallel computing and the Hadoop platform should be further investigated. Cloud computing, an efficient computation architecture that shares computing resources on the Internet, can provide different types of big data analytics services, including Platform as a Service (PaaS), Software as a Service (SaaS), and Infrastructure as a Service (IaaS) \cite{baek2015secure}. How to make full use of cloud computing resources for smart meter data analytics is an important issue. However, the security problem introduced by cloud computing should be addressed \cite{bera2015cloud}. Another high-performance computation approach is GPU computation. It can realize highly efficient parallel computation \cite{mittal2017survey}. Specific algorithms should be designed for the implementation of different GPU computation tasks.

\subsection{New Machine Learning Technologies}
Smart meter data analytics is an interdisciplinary field that involves electrical engineering and computer science, particularly machine learning. The development of machine learning has had great impacts on smart meter data analytics. The application of new machine learning technologies is an important aspect of smart meter analytics. The recently proposed clustering method in \cite{rodriguez2014clustering} has been used in \cite{wang2016clustering}; the progress in deep learning in \cite{gers2000learning} has been used in \cite{marino2016building}. When applying one machine learning technology to smart meter data analytics, the limitations of the method and the physical meaning revealed by the method should be carefully considered. For example, the size of data or samples should be considered in deep learning to avoid overfitting. 

\subsubsection{Deep Learning and Transfer Learning}
Deep learning has been applied in different industries, including smart grids. As summarized above, different deep learning techniques have been used for smart meter data analytics, which is just a start. Designing different deep learning structures for different applications is still an active research area. The lack of label data is one of the main challenges for smart meter data analytics. How to apply the knowledge learned for other objects to the research objects using transfer learning can help us fully utilize various data \cite{pan2010survey}. Many transfer learning tasks are implemented by deep learning \cite{bengio2012deep}. The combination of these two emerging machine learning techniques may have widespread applications.

\subsubsection{Online Learning and Incremental Learning}
Note that smart meter data are essentially real-time stream data. Online learning and incremental learning are varied suitably for handling these real-time stream data \cite{diethe2013online}. Many online learning techniques, such as online dictionary learning \cite{xie2014discriminative} and incremental learning machine learning techniques such as incremental clustering \cite{zhang2017incremental}, have been proposed in other areas. However, existing works on smart meter data analytics rarely use online learning or incremental learning, expect for several online anomaly detection methods.

\subsection{New Business Models in Retail Market}
Further deregulation of retail markets, integration of distributed renewable energy, and progress in information technologies will hasten various business models on the demand side. 

\subsubsection{Transactive Energy}
In a transactive energy system \cite{rahimi2012transactive} \cite{kok2016society}, the consumer-to-consumer (C2C) business model or micro electricity market can be realized, i.e., the consumer with rooftop PV becomes a prosumer and can trade electricity with other prosumers. The existing applications of smart meter data analytics are mainly studied from the perspectives of retailers, aggregators, and individual consumers. How to analyze the smart data and how much data should be analyzed in the micro electricity market to promote friendly electricity consumption and renewable energy accommodation is a new perspective in future distribution systems. 

\subsubsection{Sharing Economy}
For the distribution system with distributed renewable energy and energy storage integration, a new business model sharing economy can be introduced. The consumers can share their rooftop PV \cite{celik2017decentralized} and storage \cite{liu2017energy} with their neighborhoods. In this situation, the roles of consumers, retailers, and DSO will change when playing the game in the energy market \cite{7959054}. Other potential applications of smart meter data analytics may exist, such as changes in electricity purchasing and consumption behavior and optimal grouping strategies for sharing energy.

\subsection{Transition of Energy Systems}
\begin{figure}[!t]
  \begin{center}
  \includegraphics[width=3.3in]{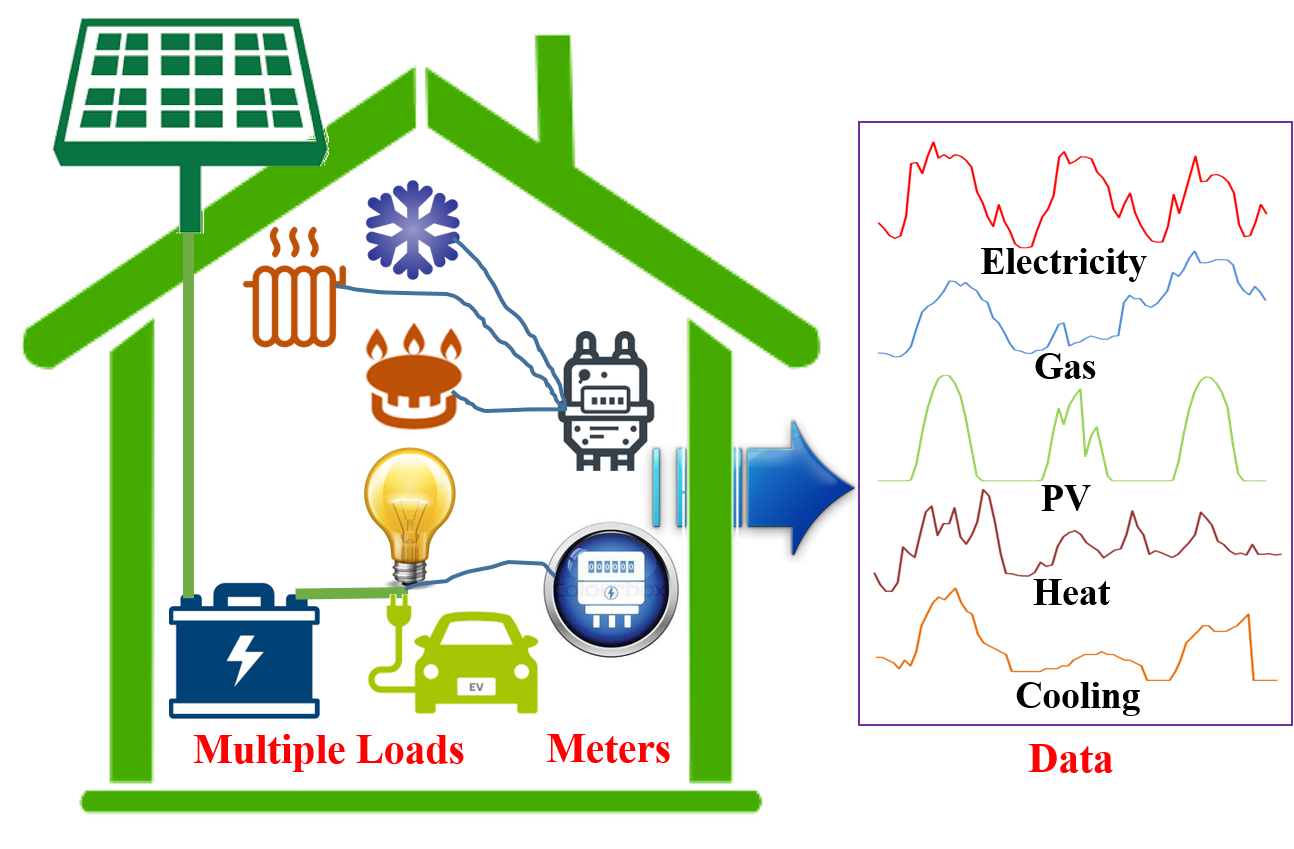}\\
  \caption{Transition of energy systems on the demand side.}\label{Transition}
  \end{center}
\end{figure}

As shown in Fig. \ref{Transition}, the integration of the distributed renewable energy and multiple energy systems is an inevitable trend in the development of smart grids. A typical smart home has multiple loads, including cooling, heat, gas, and electricity. These newcomers such as rooftop PV, energy storage, and EV also change the structure of future distribution systems.

\subsubsection{High Penetration of Renewable Energy}
High penetration of renewable energy such as behind-the-meter PV \cite{shaker2016estimating}\cite{wang2017data} will greatly change the electricity consumption behavior and will significantly influence the net load profiles. Traditional load profiling methods should be improved to consider the high penetration of renewable energy. In addition, by combining weather data, electricity price data, and net load data, the estimation of renewable energy capacity and output can be estimated. In this way, the original load profile can be recovered. Energy storage is widely used to stabilize renewable energy fluctuations. However, the charging or discharging behavior of storage, particularly the behind-the-meter storage \cite{7953569}, is difficult to model and meter. Advanced data analytical methods need to be adopted for anomaly detection, forecasting, outage management, decision making, and so forth in high renewable energy penetration environments.

\subsubsection{Multiple Energy Systems}
Multiple energy systems integrate gas, heat, and electricity systems together to boost the efficiency of the entire energy system \cite{krause2011multiple}. The consumptions for electricity, heat, cooling, and gas are coupled in the future retailer market. One smart meter can record the consumptions of these types of energy simultaneously. Smart meter data analytics is no longer limited to electricity consumption data. For example, joint load forecasting for electricity, heating, and cooling can be conducted for multiple energy systems.

\subsection{Data Privacy and Security}
As stated above, the concern regarding smart meter privacy and security is one of the main barriers to the privilege of smart meters. Many existing works on the data privacy and security issue mainly focus on the data communication architecture and physical circuits \cite{8293838}. How to study the data privacy and security from the perspective of data analytics is still limited.
\subsubsection{Data Privacy}
Analytics methods for data privacy is a new perspective except for communication architecture, such as the design of privacy-preserving clustering algorithm \cite{xing2017mutual} and PCA algorithm \cite{wei2016analysis}. A strategic battery storage charging and discharging schedule was proposed in \cite{8241403} to mask the actual electricity consumption behavior and alleviate the privacy concerns. However, several basic issues about smart meter data should be but have not been addressed: Who owns the smart meter data? How much private information can be mined from these data? Is it possible to disguise data to protect privacy and to not influence the decision making of retailers? 
\subsubsection{Data Security}
For data security, the works on cyber physical security (CPS) in the smart grid such as phasor measurement units (PMU) and supervisory control and data acquisition (SCADA) data attack have been widely studied \cite{yan2012survey}. However, different types of cyber attacks for electricity consumption data such as NTL should be further studied \cite{zhang2017achieving}.

\section{Conclusions}
\label{conclusion}
In this paper, we have provided a comprehensive review of smart meter data analytics in retail markets, including the applications in load forecasting, abnormal detection, consumer segmentation, and demand response. The latest developments in this area have been summarized and discussed.  In addition, we have proposed future research directions from the prospective big data issue, developments of machine learning, novel business model, energy system transition, and data privacy and security. Smart meter data analytics is still an emerging and promising research area. We hope that this review can provide readers a complete picture and deep insights into this area.

\bibliographystyle{IEEEtran}
\bibliography{IEEEabrv,Reference}

\begin{IEEEbiography}[{\includegraphics[width=1in,height=1.25in,clip,keepaspectratio]{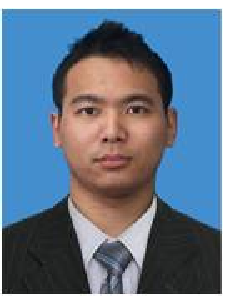}}]{Yi Wang}
(S'14) received the B.S. degree from the Department of Electrical Engineering in Huazhong University of Science and Technology (HUST), Wuhan, China, in 2014.\\
He is currently pursuing Ph.D. degree in Tsinghua University. He is also a visiting student researcher at the University of Washington, Seattle, WA, USA. His research interests include data analytics in smart grid and multiple energy systems.
\end{IEEEbiography}

\begin{IEEEbiography}[{\includegraphics[width=1in,height=1.25in,clip,keepaspectratio]{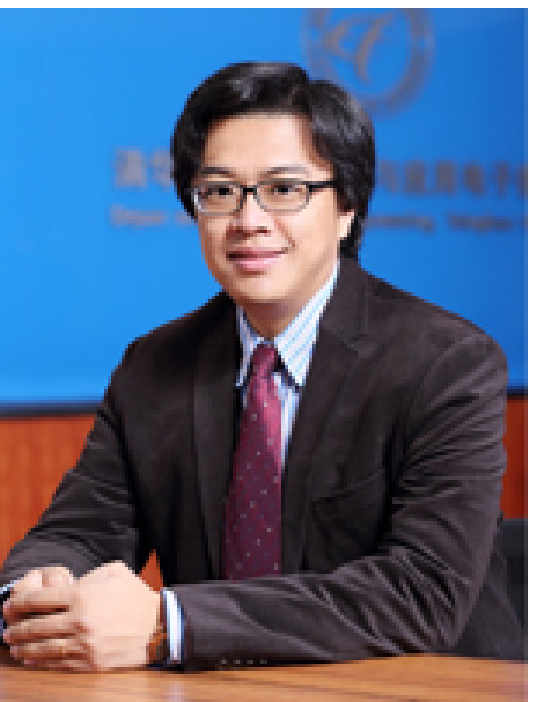}}]{Qixin Chen}
(M'10-SM'15) received the Ph.D. degree from the Department of Electrical Engineering, Tsinghua University, Beijing, China, in 2010.\\
He is currently an Associate Professor at Tsinghua University. His research interests include electricity markets, power system economics and optimization, low-carbon electricity, and power generation expansion planning.
\end{IEEEbiography}

\begin{IEEEbiography}[{\includegraphics[width=1in,height=1.25in,clip,keepaspectratio]{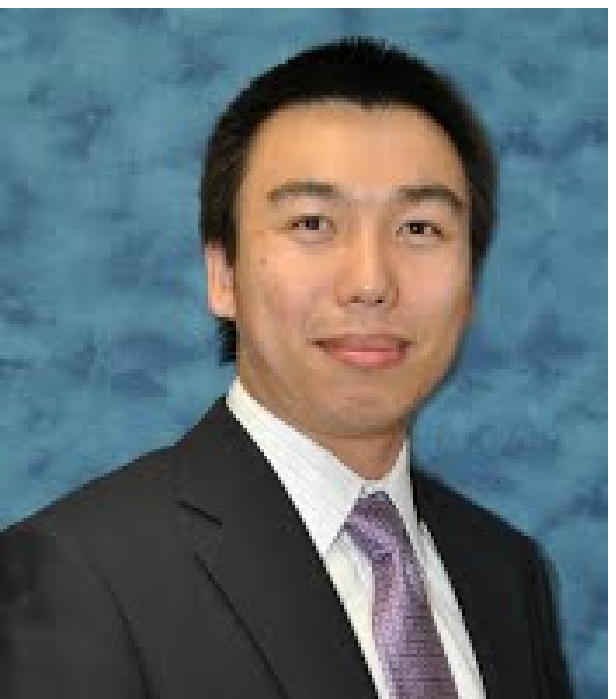}}]{Tao Hong}
received the B.Eng. degree in automation from Tsinghua University, Beijing, China, in 2005, and the Ph.D. degree in operation research and electrical engineering from North Carolina State University, Raleigh, NC, USA, in 2010. \\
He is the Director of the Big Data Energy Analytics Laboratory (BigDEAL), and an Associate Professor of System Engineering and Engineering Management at the University of North Carolina at Charlotte, Charlotte, NC. Dr. Hong is the Founding Chair of the IEEE Working Group on Energy Forecasting and the General Chair of the Global Energy Forecasting Competition.
\end{IEEEbiography}

\begin{IEEEbiography}[{\includegraphics[width=1in,height=1.25in,clip,keepaspectratio]{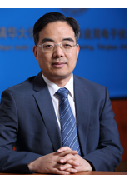}}]{Chongqing Kang}
(M'01-SM'08-F'17) received the Ph.D. degree from the Department of Electrical Engineering in Tsinghua University, Beijing, China, in 1997. \\
He is currently a Professor in Tsinghua University. His research interests include power system planning, power system operation, renewable energy, low carbon electricity technology and load forecasting.
\end{IEEEbiography}

\ifCLASSOPTIONcaptionsoff
\fi
\end{document}